\documentclass[preprint,pre,superscriptaddress]{revtex4}
\usepackage{graphicx}
\usepackage{lipsum}
\usepackage{float}
\usepackage{amsmath}
\usepackage{amssymb}
\usepackage{mathtools}

\begin{document}

\title{
The distribution of shortest path lengths in
subcritical Erd{\H o}s-R\'enyi networks 
}

\author{Eytan Katzav}
\affiliation{
Racah Institute of Physics, 
The Hebrew University, 
Jerusalem 91904, Israel}

\author{Ofer Biham}
\affiliation{
Racah Institute of Physics, 
The Hebrew University, 
Jerusalem 91904, Israel}

\author{Alexander K. Hartmann}
\affiliation{
Institute of Physics, 
Carl von Ossietzky University, 
26111 Oldenburg, Germany
}

\begin{abstract}

Networks that are fragmented into small disconnected 
components are prevalent in a large variety of systems. 
These include the secure communication networks
of commercial enterprises, government agencies and
illicit organizations, as well as networks that suffered
multiple failures, attacks or epidemics.  
The structural and statistical properties of such networks 
resemble those of subcritical random networks, 
which consist of finite components, 
whose sizes are non-extensive.
Surprisingly, such networks do not exhibit the small-world 
property that is typical in supercritical random networks,
where the mean distance between pairs of nodes
scales logarithmically with the network size.
Unlike supercritical networks whose structure has been 
studied extensively, subcritical networks have attracted
relatively little attention.
A special feature of these networks is that the 
statistical and geometric properties vary between
different components
and depend on their sizes and topologies.
The overall statistics of the network can be obtained by a summation
over all the components with suitable weights.
We use a topological expansion to
perform a systematic analysis of the 
degree distribution and the
distribution of shortest path lengths (DSPL) on
components of given sizes and topologies
in subcritical
Erd{\H o}s-R\'enyi (ER) networks.
From this expansion we
obtain an exact analytical expression for
the DSPL of the entire subcritical network,
in the asymptotic limit. 
The DSPL, which accounts for all the pairs of nodes that
reside on the same finite component (FC),
is found to follow a geometric
distribution of the form
$P_{\rm FC}(L=\ell | L < \infty) = (1-c) c^{\ell - 1}$,
where $c<1$ is the mean degree.
Using computer simulations we calculate
the DSPL in subcritical ER networks of increasing sizes
and confirm the convergence 
to this asymptotic result.
We also obtain exact asymptotic results for the mean distance,
$\langle L \rangle_{\rm FC}$, and for the standard deviation of the 
DSPL, $\sigma_{L,{\rm FC}}$, and show that the simulation results
converge to these asymptotic results.
Using the duality relations between subcritical and supercritical ER networks, 
we obtain the DSPL on
the non-giant components of ER networks 
above the percolation transition.

\end{abstract}


\maketitle

\section{Introduction}

Network models provide a useful framework for the analysis of
a large variety of systems that consist of interacting objects
\cite{Newman2010,Havlin2010,Estrada2011}.
In these models, the objects are represented by nodes and
the interactions between them are described by edges.
A pair of nodes, $i$ and $j$, may be connected via many
different paths. The shortest among these paths are of
particular importance because they provide the fastest 
and often the strongest interaction.
Broadly speaking, one
can distinguish between two major classes of networks:
supercritical networks, which are tightly connected, and  
subcritical networks, which are loosely connected. 
Supercritical networks form a giant component
that encompasses a macroscopic fraction of all the nodes,
while the typical distance between pairs of nodes on
the giant component scales logarithmically with the network size.
Examples of such networks are the
world-wide-web, social networks, and infrastructure networks
of transportation, telephone, internet and electricity.
In contrast, subcritical networks are fragmented into small
components that do not scale with the overall network size.
Examples of fragmented networks include
secure networks with controlled access,
such as the communication networks 
of commercial enterprises, government agencies
and illicit organizations
\cite{Duijn2014}.
Other examples include networks that suffered multiple failures, large scale attacks or epidemics,
in which the remaining functional or uninfected nodes form small, isolated components
\cite{Shao2008,Shao2009}.
In spite of their prevalence, fragmented networks are of low visibility 
and have not attracted nearly as much attention as 
supercritical networks.

Random networks of the 
Erd{\H o}s-R\'enyi (ER) type
\cite{Erdos1959,Erdos1960,Erdos1961} 
are the simplest class of random networks and
are used as a benchmark for the study of structure
and dynamics in complex networks
\cite{Bollobas2001}.
The ER network ensemble
is a maximum entropy ensemble, 
under the condition
that the mean degree
$\langle K \rangle = c$
is fixed.
It is a special case of a broader class of random uncorrelated networks,
referred to as configuration model networks
\cite{Newman2001,Newman2010,Klein2012}.
In an ER network of $N$ nodes, 
each pair of nodes
is independently connected with probability $p$,
such that the mean degree is
$c=(N-1)p$.
The ensemble of such networks is denoted by ER[$N,p$].
The degree distribution of these networks follows a
Poisson distribution of the form

\begin{equation}
\pi(K=k) = \frac{e^{-c} c^{k}}{k!}.
\label{eq:poisson}
\end{equation}

\noindent
ER networks exhibit a percolation transition 
at $c=1$ such that for $c>1$
there is a giant component,
while for $0 < c < 1$ the network consists of small, isolated tree components
\cite{Bollobas2001,Durrett2007}.
The probability that a random node in the network resides on the giant
component is denoted by $g=g(c)$.
Clearly, below the percolation transition
$g=0$.
Above the transition
\cite{Bollobas2001}

\begin{equation}
g(c) = 1 + \frac{\mathcal{W}(-c e^{-c})}{c},
\label{eq:g(c)}
\end{equation}

\noindent
where $\mathcal{W}(x)$ is the Lambert $W$ function
\cite{Olver2010}.
For networks in the range $1 < c < \ln N$, the probability $g$ satisfies
$0 < g < 1$, namely the giant and finite components coexist,
while for $c > \ln N$ the giant component encompasses the whole network
and $g=1$.

To characterize the paths connecting
random pairs of nodes, 
measures such as the mean distance and the diameter
were studied 
\cite{Chung2002,Hofstad2005,Bollobas2007,Newman2001,Fronczak2004,Hartmann2017}.
For supercritical ER networks, it was shown that the mean distance,
$\langle L \rangle$,
scales like
$\langle L \rangle \sim \ln N / \ln c$,
in agreement with rigorous results,
showing that percolating random networks are small-world networks
\cite{Chung2002,Bollobas2007}.
For subcritical ER networks it was recently shown that the distribution of
diameters over an ensemble of networks follows a Gumbel distribution
of extreme values
\cite{Hartmann2017}.
This is due to the fact that in subcritical networks the diameter is obtained by maximizing
the distances over all the small components.
For supercritical networks,
the entire distribution of shortest path lengths (DSPL)
was calculated using various approximation techniques
\cite{Dorogotsev2003,Blondel2007,Hofstad2007,Esker2008,Shao2008,Shao2009,Katzav2015,Nitzan2016,Melnik2016}.
However, the DSPL of subcritical networks has not been studied.

The DSPL provides
a natural platform for 
the study of dynamical processes on networks,
such as 
diffusive processes,
epidemic spreading,
critical phenomena,
synchronization,
information propagation and communication
\cite{Havlin2010,Newman2010,Estrada2011,Satorras2015}.
Thermal and dynamical processes on networks
resemble those of systems with long range interactions
\cite{Bouchet2010}
in the sense that extensivity is broken and 
standard statistical physics techniques
do not apply.
Therefore, it is important to develop theoretical approaches 
that take into account the topological and geometrical properties
of complex networks.
In fact, the DSPL provides exact 
solutions for various dynamical problems 
on networks.
In the context of traffic flow on networks,
the DSPL provides the distribution of transit times 
between all pairs of nodes, 
in the limit of low traffic load
\cite{Barrat2008}.
In the context of search processes,
the DSPL
determines the order in which nodes
are explored in the breadth-first search protocol
\cite{Barrat2008}.
In the context of epidemic spreading,
the DSPL captures the temporal evolution of
the susceptible-infected epidemic,
in the limit of high infection rate
\cite{Satorras2015}.
In the context of network attacks, 
the DSPL describes a generic class of violent local attacks,
which spreads throughout the network
\cite{Shao2015}.
It is also used as a measure that quantifies the structural
dissimilarities between different networks
\cite{Schieber2017}.

The DSPL provides a useful characterization of empirical networks.
For example, the DSPL of the protein network in Drosophila melanogaster
was compared to the DSPL 
of a corresponding randomized network
\cite{Giot2003}.
It was shown that
proteins in this network are significantly farther
away from each other than in the randomized network,
providing useful biological insight. 
In the context of brain research, it was found that the DSPL
and the distribution of shortest cycle lengths
\cite{Bonneau2017}
determine the periods of oscillations in the activity of
neural circuits
\cite{Goldental2015,Goldental2017}.
In essence, shortest paths and shortest cycles control the
most important feedback mechanisms in these circuits, setting
the characteristic time scales at which oscillations are sustained. 

As mentioned above, 
in the asymptotic limit,
$N \rightarrow \infty$,
ER networks exhibit a percolation transition 
at $c=1$. 
For $c<1$, an ER network consists of finite components (FC),
which are non-extensive in the network size, while for 
$c>1$ a giant component (GC) 
is formed, which includes a finite fraction of the nodes
in the network
\cite{Bollobas2001}.
When two nodes,
$i$ and $j$, 
reside on the same 
component, 
the distance, 
$\ell_{ij}$, 
between them
is defined as the length of the shortest path 
that connects them.
In the networks studied here, whose edges do not carry distance labels,
the length of a path is given by the number of edges along the path.
When $i$ and $j$ reside on different components,
there is no path connecting them
and we define the distance between them to be
$\ell_{ij} \equiv \infty$.
We denote the probability distribution
$P_{\rm FC}(L=\ell)$
as the DSPL over all 
$\binom{N}{2}$
pairs of nodes in a subcritical ER network.
The probability that two randomly selected
nodes reside on the same component, and
thus are at a finite distance from each other,
is denoted by
$P_{\rm FC}(L<\infty) = 1 - P_{\rm FC}(L=\infty)$. 

Here we focus on the conditional DSPL between pairs 
of nodes that reside on the same component,
denoted by
$P_{\rm FC}(L=\ell | L<\infty)$,
where
$\ell=1,2,\dots,N-1$.
The conditional DSPL satisfies

\begin{equation}
P_{\rm FC}(L=\ell | L<\infty)=\frac{P_{\rm FC}(L=\ell)}{P_{\rm FC}(L<\infty)}.
\end{equation}

In this paper we use a topological expansion to perform a systematic analysis of the
degree distribution and the
DSPL on finite tree components of all sizes and topologies
in subcritical ER networks.
We find that in the asymptotic limit the DSPL
is given by a geometric
distribution of the form
$P_{\rm FC}(L=\ell | L<\infty) = (1-c) c^{\ell - 1}$,
where $c<1$.
Using computer simulations we calculate
the DSPL in subcritical ER networks of increasing sizes
and confirm the convergence 
to this asymptotic result.
We also show that the mean
distance between pairs of 
nodes that reside on the same component is given by
$\langle L \rangle_{\rm FC} = 1/(1-c)$.
The average size of the tree components (obtained by random sampling of trees) is
$\langle S \rangle_{\rm FC}=2/(2-c)$.
However, the average tree component size, obtained by 
random sampling of nodes, is 
$\langle \widetilde S \rangle_{\rm FC} = 1/(1-c)$.
Thus, the mean distance turns out to scale linearly with
the average tree component size on which a random node resides.
This is in contrast to
supercritical random networks,
in which the mean distance
scales logarithmically or even sub-logarithmically with the network size
\cite{Chung2002,Hofstad2005,Bollobas2007,Cohen2003}.
Using duality relations
connecting the non-giant components of supercritical ER networks to
the corresponding subcritical ER networks
\cite{Bollobas2001,Molloy1995,Molloy1998},
we obtain the DSPL of 
the non-giant components of the ER network 
above the percolation transition.

The paper is organized as follows.
In Sec. II we describe recent results for the DSPL of supercritical networks.
In Sec. III we review some fundamental properties of subcritical
ER networks, which are used in the analysis below.
In Sec. IV we present the topological expansion.
In Sec. V we apply the topological expansion to calculate
the degree distribution of subcritical ER networks.
In Sec. VI we calculate the mean and variance of the degree distribution.
In Sec. VII we apply the topological expansion to calculate the DSPL 
of subcritical ER networks.
In Sec. VIII we calculate the mean and variance of the DSPL.
The results are discussed in Sec. IX and summarized in Sec. X.
In Appendix A we present the calculation of the component size distribution, $P_{\rm FC}(S=s)$,
in subcritical ER networks, which is used in the topological expansion.
In Appendix B we present the calculation of the probability, 
$P_{\rm FC}(L<\infty)$,
that two random nodes in a subcritical ER network reside on the same component.

\section{The DSPL of supercritical Erd{\H o}s-R\'enyi networks}

Consider a pair of random nodes, $i$ and $j$, in a supercritical ER
network of $N$ nodes. 
The probability that both of them reside on the giant component is $g^2$.
The probability that one of them resides on the giant component and the other
resides on one of the finite components is $2g(1-g)$, while the probability that
both reside on finite components is $(1-g)^2$.
In case that both nodes
reside on the giant component, they are
connected to each other by at least one path.
Therefore, the distance between them is finite. 
In case that one node resides on the giant component while the
other node resides on one of the finite components, the distance between them is
$\ell_{ij}=\infty$.
In case that both nodes reside on finite components, a path
between them exists only in the low probability case that they reside on
the same component.
The finite components are trees and therefore the shortest path between
any pair of nodes is unique.

The DSPL of a supercritical ER network 
can be expressed by

\begin{equation}
P(L=\ell) 
= 
g^2 P_{\rm GC}(L=\ell)
+ (1-g)^2 P_{\rm FC}(L = \ell),
\label{eq:fullDSPL}
\end{equation}

\noindent
where the first term accounts for the DSPL between pairs
of nodes that reside on the giant component and the
second term accounts for pairs of nodes that reside on
finite components.
This form is particularly useful in the range of 
$1 < c < \ln N$, in which the giant component and the
finite components coexist.
The probability that there is no path connecting a random pair of  
nodes is given by

\begin{equation}
P(L = \infty) =
2 g (1-g)
+(1-g)^2 P_{\rm FC}(L=\infty).
\label{eq:LinfGF}
\end{equation}

\noindent
The first term in Eq. (\ref{eq:LinfGF})
accounts for the probability that one node resides
on the giant component while the other resides on one of the finite components.
The second term accounts for the probability that the two nodes reside on two
different finite components and are thus not connected by a path.
In order to obtain accurate results for the DSPL of an ER network in this regime,
one needs to calculate both the DSPL of the giant component,
$P_{\rm GC}(L=\ell)$, and the DSPL of the finite components,
$P_{\rm FC}(L=\ell) = P_{\rm FC}(L=\ell | L < \infty) P_{\rm FC}(L<\infty)$.
The giant component of an ER network with $1 < c < \ln N$ is
a more complicated geometrical object than the whole network. 
Its degree distribution
deviates from the Poisson distribution and it exhibits degree-degree
correlations.
The degree distribution and degree-degree correlations in the 
giant component of supercritical ER networks with $1 < c < \ln N$ 
were recently studied
\cite{Tishby2018}.
Using these results, the DSPL of the giant component was calculated 
\cite{Tishby2018}.

%
%
For $c > \ln N$ the network consists of a single connected component
and the DSPL of the whole network can be calculated using the recursion
equations presented in Refs. 
\cite{Katzav2015,Nitzan2016}.
In this approach one denotes the conditional probability,
$P(L>\ell | L>\ell-1)$,
that the distance between
a random pair of nodes, $i$ and $j$, is larger than $\ell$, 
under the condition that it is
larger than $\ell-1$.
A path 
of length $\ell$
from node $i$ to node $j$ can be decomposed 
into a single edge connecting node $i$ and node 
$r \in \partial_i$
(where $\partial_i$ is the set of all nodes directly connected to $i$),
and a shorter path of length 
$\ell-1$ connecting $r$ and $j$.
Thus, the existence of a path of length $\ell$
between nodes $i$ and $j$
can be ruled out if there is no path of length
$\ell-1$ between any of the nodes 
$r \in \partial_i$,
and $j$.
For sufficiently large networks,
the argument presented above translates into
the recursion equation
\cite{Nitzan2016}

\begin{equation}
P(L > \ell | L>\ell-1) =
G_0[P(L>\ell-1 | L> \ell-2)],
\label{eq:P_rec2}
\end{equation}

\noindent
where  

\begin{equation}
G_0(x) = \sum_{k=0}^{\infty}  x^{k}   \pi(K=k),
\label{eq:G0}
\end{equation}

\noindent
is the generating function of the Poisson distribution.
The case of $\ell=1$,
used as the initial condition for the recursion equations
is given by
$P(L>1 | L>0) = 1 - c/(N-1)$.
The recursion equations provide a good approximation for the DSPL
of supercritical ER networks
\cite{Katzav2015,Nitzan2016,Melnik2016},
for values of $c$ that are sufficiently far above the percolation threshold.
However, no exact result for the DSPL of supercritical ER networks is known.
Interestingly, for random regular graphs, in which all the nodes are of the
same degree, $k=c \ge 3$, there is an exact result for the DSPL, which can be expressed by
a Gompertz distribution 
\cite{Gompertz1825}
of the form
\cite{Hofstad2005,Nitzan2016}

\begin{equation}
P(L > \ell) = \exp \left[{-\beta (e^{b \ell} - 1)} \right],
\end{equation}

\noindent
where 
$\beta=c/[N(c-2)]$ and $b = \ln (c-1)$.
For a supercritical ER network with mean degree $c$,
which is sufficiently far above the percolation threshold,
the DSPL is qualitatively similar to the
DSPL of a random regular graph with degree 
$\lfloor c+1/2 \rfloor$. 
Here, $\lfloor x \rfloor$ is
the integer part of $x$ and thus 
$\lfloor x+1/2 \rfloor$ is the
nearest integer to $x$.
Unlike random regular graphs in which all the nodes are of the same degree,
in supercritical ER networks, which follow the Poisson degree distribution,
the shortest path length between 
a pair of nodes is correlated with the degrees of these nodes.
The correlation is negative, namely
nodes of high degrees tend to be closer to each other than nodes of low degrees.
Another simplifying feature of random regular graphs with $c \ge 3$ is that the giant component
encompasses the entire network ($g=1$) and thus all pairs of nodes are connected by
finite paths.
Since ER networks with $1 < c < \ln N$ consist of a combination of a giant component and finite
components, the DSPL exhibits a non-zero asymptotic tail and its calculation is more difficult.

The DSPL on the finite components in supercritical ER networks
is a sub-leading term in the overall DSPL, which involves a fraction of
$(1-g)^2 P_{\rm FC}(L<\infty)$
from the $\binom{N}{2}$ pairs of nodes in the network.
The factor of $(1-g)^2$ accounts for the fraction of pairs
that reside on the finite components, while the fraction of those pairs
that reside on the same component is given by
$P_{\rm FC}(L<\infty)$.
Except for the vicinity of the percolation transition, which occurs at $c=1$,
this amounts to a small fraction of all pairs of nodes.

In the asymptotic limit,
ER networks exhibit duality with respect to the percolation threshold.
In a supercritical ER
network of $N$ nodes 
the fraction of nodes that belong to the giant component is 
$g$
[Eq. (\ref{eq:g(c)})],
while the fraction of nodes that belong to the finite components is
$f=1-g$.
Thus, the subcritical network that consists of the finite components
is of size $N^{\prime} = N f$. 
This network 
is an ER network 
whose mean degree is
$c^{\prime} = c f$,
where
$c^{\prime}<1$.
It means that the DSPL of the finite components
of a supercritical ER network can be obtained from the analysis
of its dual subcritical network
\cite{Tishby2018}.

\section{Subcritical Erd{\H o}s-R\'enyi networks}

In the analysis presented below we use
the fact that the components that appear in subcritical ER networks
are almost surely trees,
namely the expected number of cycles is non-extensive
\cite{Bollobas2001}.
In Appendix A we show that the expected number of tree components
in a subcritical ER network of $N$ nodes is

\begin{equation}
N_{T}(c) = \left( 1 - \frac{c}{2} \right) N,
\end{equation}

\noindent
and the distribution of tree sizes is
\cite{Bollobas2001,Durrett2007}

\begin{equation}
P_{\rm FC}(S=s) = \frac{2 s^{s-2} c^{s-1} e^{-cs}}{(2-c)s!}.
\end{equation}

\begin{figure}
\begin{center}
\includegraphics[width=13cm]{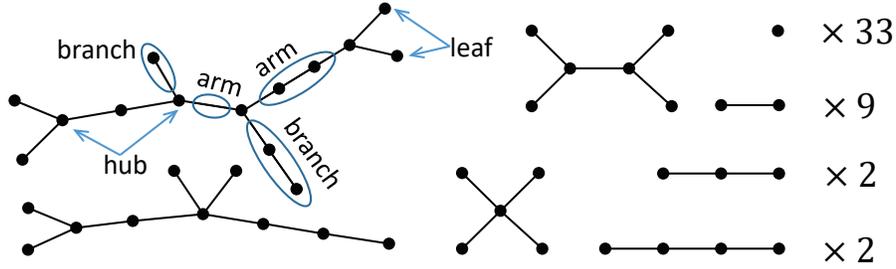}
\end{center}
\caption{
(Color online)
Illustration of the structure of one instance of a subcritical ER network 
of $N=100$ nodes and $c=0.9$.
It consists of 33 isolated nodes, 9 dimers, two chains of three nodes,
two chains of four nodes, one tree with a single hub and four branches,
one tree with two hubs and two larger trees of 10 and 14 nodes.
}
\label{fig:1}
\end{figure}

\noindent
In these trees we define all the nodes of degree 
$k \ge 3$ as {\it hubs} and
all the nodes of degree $k=1$ as {\it leaves}.
Linear chains of nodes that have a hub on one side 
and a leaf on the other side are referred to as {\it branches}, 
while chains that have hubs on both sides
are referred to as {\it arms}. 
In Fig. \ref{fig:1} 
we illustrate the structure of an ER network
of size $N=100$ and $c=0.9$. 
It consists of 33 isolated nodes, 9 dimers, two chains of three nodes,
two chains of four nodes, one tree with a single hub and four branches,
one tree with two hubs and two larger trees of 10 and 14 nodes.

Selecting two random nodes in a subcritical ER network, 
the probability that they reside on the 
same component is denoted by $P_{\rm FC}(L < \infty)$.
In Appendix B we show that

\begin{equation}
P_{\rm FC}(L < \infty) = \frac{c}{(1-c)N}. 
\label{eq:L<infty}
\end{equation}

\noindent
Using this result 
and the fact that the first two terms of 
$P_{\rm FC}(L=\ell)$
are known exactly,
namely
$P_{\rm FC}(L=1)=p$
and
$P_{\rm FC}(L=2) = (1-p)[1-(1-p^2)^{N-2}]$
\cite{Katzav2015},
we obtain that
$P_{\rm FC}(L=1 | L<\infty)=1-c$
and
$P_{\rm FC}(L=2 | L<\infty)=c(1-c)$.

In the next Section we introduce the topological expansion method.
In this approach, for each component size, $s$, 
we identify all the possible tree topologies supported by
$s$ nodes, starting from the linear chain, which
does not include any hubs, followed by single-hub topologies,
double-hub topologies and higher order topologies, which
include multiple-hubs.
For each tree topology, we calculate
its relative weight among all possible tree topologies 
of the same size.
A special property of
tree topologies is that each pair of nodes is connected by a single path. 
Therefore, in subcritical ER networks
the shortest path between any pair of nodes is, in fact, the only path between them.
Using this property we
calculate the DSPL for 
each tree topology, and use the weights to obtain the 
DSPL over all the components that consist of up to $s$ nodes.

\section{The topological expansion}

Consider a tree that includes $h$ hubs.
Embedded in this tree, there is a backbone tree, which
consists only of the $h$ hubs and the $h-1$ arms that connect them. 
The structure of the backbone tree is described by its adjacency matrix, $A$.
This is a symmetric $h \times h$ matrix, 
in which $A_{ij}=1$ if hubs $i$ and $j$ are connected by an arm,
and $0$ otherwise.
The degrees of the hubs in the backbone tree are denoted by
the vector

\begin{equation}
\vec a = (a_1,a_2,\dots,a_h),
\end{equation}

\noindent
where 

\begin{equation}
a_i = \sum_{j=1}^h A_{ij}.
\end{equation}

\noindent
The structure of the branches is described by
the vector

\begin{equation}
\vec b = (b_1,b_2,\dots,b_h),
\end{equation}

\noindent
where $b_i$ is the number of branches connected to the $i^{\rm th}$ hub.
The total number of branches in a tree is given by

\begin{equation}
b = \sum_{i=1}^h b_i.
\end{equation}

\noindent
The topology of a tree with $h$ hubs is fully described by the
adjacency matrix, $A$, of its backbone tree
and its branch vector $\vec b$.
We denote such tree topology by

\begin{equation}
\tau = (h,A,\vec b).
\end{equation}

\noindent
In this classification, the linear chain structure is denoted by
$\tau = (0,\cdot,2)$. 
It has no nodes and thus $h=0$. The matrix $A$ is not defined and replaced
by the "$\cdot$" sign. 
The linear chain has two leaf nodes and it is thus considered as a tree with two branches.
A tree that includes a single hub with $b \ge 3$ branches is denoted by
$\tau = (1,0,b)$.
Here, the matrix $A=0$ is a scalar.
A tree that includes two hubs
with a branch vector 
$\vec b = (b_1,b_2)$,
is denoted by 
$\tau = (2,A,\vec b)$,
where $A$ is a $2 \times 2$ matrix
of the form

\begin{equation}
A =
\left( 
\begin{array}{cc}
0 & 1  \\
1 & 0 
\end{array} 
\right).
\label{eq:2by2}
\end{equation}

\noindent
A tree that consists of a chain of three hubs,
with a branch vector 
$\vec b = (b_1,b_2,b_3)$,
is denoted by 
$\tau = (3,A,\vec b)$,
where $A$ is a $3 \times 3$ matrix
of the form

\begin{equation}
A=
\left( 
\begin{array}{ccc}
0 & 1 & 0 \\
1 & 0 & 1 \\
0 & 1 & 0
\end{array} 
\right).
\end{equation}

\noindent
A tree that includes four hubs,
consisting of one central hub surrounded by three peripheral hubs
is denoted by
$\tau =(4,A,\vec b)$
where

\begin{equation}
A =
\left( 
\begin{array}{cccc}
0 & 1 & 1 & 1\\
1 & 0 & 0 & 0\\
1 & 0 & 0 & 0 \\
1 & 0 & 0 & 0
\end{array} 
\right).
\end{equation}

\noindent
and
$\vec b = (b_1,b_2,b_3,b_4)$.

\begin{figure}
\begin{center}
\includegraphics[width=12cm]{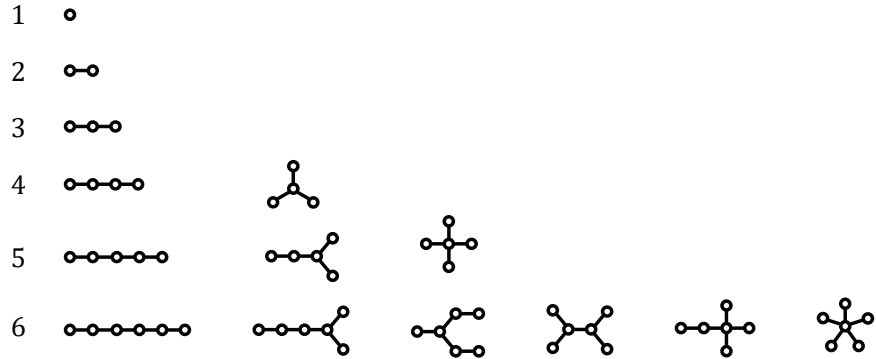}
\end{center}
\caption{
A list of all the possible backbone tree topologies that consist of up to six hubs
($h=1,2,\dots,6$).
The linear chain topology appears for all values of $h$.
For $h \le 3$ it is the only topology, while for $h \ge 4$
more complex topologies appear and their number quickly
increases.
}
\label{fig:2}
\end{figure}

In Fig. \ref{fig:2} we present all the possible topologies of the backbone
tree that can be obtained with up to six hubs.
The linear chain topology exists for all values of $h$.
For $h \le 3$ it is the only topology, while for $h \ge 4$ additional
topologies appear and their number quickly increases.
More specifically, for
$h=1$ the backbone tree is a single hub,
for $h=2$ it is a dimer and for
$h=3$ it is a linear chain of three hubs.
For $h=4$ there are two possible tree topologies: a linear chain of hubs
and a tree that consists of a central hub surrounded by three peripheral hubs.
For $h=5$ there are three possible topologies while for $h=6$ there are
six possible topologies.

The topological expansion is performed
such that the $s^{\rm th}$ 
order consists of all possible
tree topologies that can be 
assembled from $s$ nodes.
Since each branch consists of at least one node,
the number of branches in a tree that consists of $s$ nodes
and includes $h$ hubs must satisfy

\begin{equation}
b \le  s - h.
\label{eq:s1}
\end{equation}

\noindent
Unlike the branches, 
each arm may either consist of a single edge between the adjacent
hubs or include one or more intermediate nodes.
The number of arms connecting the $h$ hubs in the tree is $h-1$.
The degree of each hub is given by the sum of the number of branches
and the number of arms connected to it. 
While each branch is connected
to only one hub, each arm is connected to two hubs, one on each side.
Recalling that the degree of each hub is $k \ge 3$ we find that
$2(h-1)+b \ge 3h$.
Thus, the number of branches in a tree that includes $h$ 
hubs must satisfy

\begin{equation}
b \ge h+2.
\label{eq:b}
\end{equation}

\noindent
Combining Eqs. (\ref{eq:s1}) and (\ref{eq:b})
we obtain a condition on the minimal tree size that may include
$h$ hubs. It takes the form

\begin{equation}
s \ge 2h+2.
\label{eq:s2}
\end{equation}

\noindent
We thus obtain a classification of all tree
structures that can be assembled from $s$ nodes,
where $s \ge 2$.
For  $s=2,3$ the linear chain is the only possible topology.
Higher order topologies, 
which exist for $s \ge 4$, include at least one hub. 
In a tree of size $s \ge 4$, 
the number of hubs may take values in the range

\begin{equation}
h = 1,2,\dots, \left\lfloor \frac{s}{2}-1 \right\rfloor.
\end{equation}

\noindent
For each choice of $h$, the number of branches 
may take any value in the range

\begin{equation}
b = h+2, h+3,\dots,s-h.
\label{eq:bvalues}
\end{equation}

\begin{figure}
\begin{center}
\includegraphics[width=9cm]{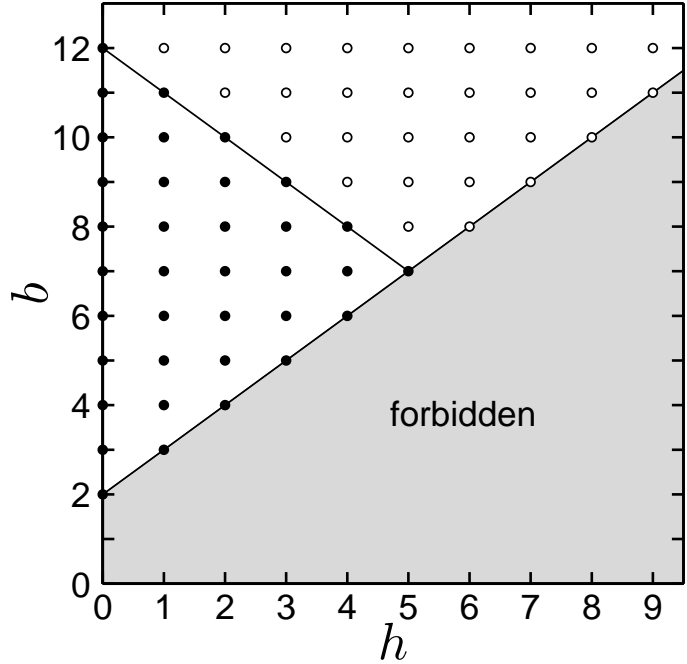}
\end{center}
\caption{
Illustration of the range of possible values of $b$ (the number of branches)
in a tree of $h$ hubs, which consists of $s$ nodes.
This range is bounded from below by $b = h+2$, due to a topological
constraint, regardless of $s$ (ascending straight line).
For a network that consists of $s$ nodes, is is bounded from
above by $b = s-h$ (descending straight line).
The two lines intersect at $(h,2h+2)$.
Combinations of $(h,b)$ that are possible in a tree of $s=12$ nodes 
are marked by full circles, while combination that exist only in larger
trees are marked by empty circles.
Each backbone tree can be represented by its adjacency matrix $A$,
which is an $h \times h$ matrix.
The topology of a complete tree is denoted by $\tau = (h,A,\vec b)$,
where $\vec b = (b_1,\dots,b_h)$
accounts for a specific division of the $b$ branches between the
$h$ hubs.
}
\label{fig:3}
\end{figure}

In Fig. \ref{fig:3} we illustrate
the possible values of $b$ 
in a tree of $h$ hubs, which consists of $s$ nodes,
given by Eq. (\ref{eq:bvalues}), i.e., the
range is bounded from below by $b = h+2$
and from
above by $b = s-h$.
Combinations of $(h,b)$ that are possible in a tree of $s=12$ nodes 
are marked by full circles, while combinations that exist only in larger
trees are marked by empty circles.

The number of non-isomorphic tree topologies, $n(h)$, which can be assembled
from $h$ nodes quickly increases as a function of $h$. 
For example, the values of $n(h)$ for $h=1,2,\dots,13$ are
1, 1, 1, 2, 3, 6, 11, 23, 47, 106, 235, 551 and 1301, respectively
\cite{Steinbach1990}.
An efficient algorithm
for generating all the tree topologies
that can be assembled from
$h$ nodes, is presented in Refs.
\cite{Wright1986,Beyer1980}.
A list of all possible tree topologies up to $h=13$ is presented in Ref.
\cite{Steinbach1990}.
Other web resources include enumeration of such tree topologies
up to $h=20$
\cite{Royle}.

Each one of the possible topologies of the backbone tree is represented by its adjacency matrix, $A$,
which is an $h \times h$ matrix.
The degrees of the hubs in the backbone tree are given by 
$\vec a = (a_1,a_2,\dots,a_h)$.
Since the degrees of the hubs, which are given by $k_i=a_i+b_i$,
must satisfy the condition $k_i \ge 3$, the components of the 
branch vector, $\vec b = (b_1,b_2,\dots,b_h)$ satisfy
the condition

\begin{equation}
b_i \ge (3-a_i) \theta(3-a_i),
\end{equation}

\noindent
where
$\theta(x)$ is the Heaviside function, which satisfies
$\theta(x) =1$ for $x > 0$ and $\theta(x)=0$ for $x \le 0$.
The number of branches required to satisfy this condition is 
$h+2$. In case that $b>h+2$, the remaining $b-h-2$ branches
can be divided in many different ways between the $h$ hubs.
The number of possible
partitions of $x$ identical objects among $y$ distinguishable boxes is given by 
the multiset coefficient
\cite{Stanley2011}

\begin{equation}
\left(\hspace{-0.05in}\binom{y}{x}\hspace{-0.05in}\right) = \binom{x+y-1}{x} = \binom{x+y-1}{y-1}.
\label{eq:partitions}
\end{equation}

\noindent
Therefore, the number of different tree topologies 
that can be obtained
from a single topology of a backbone tree of $h$ hubs is

\begin{equation}
\left(\hspace{-0.05in}\binom{h}{b-h-2}\hspace{-0.05in}\right)  = \binom{b-3}{h-1}.
\end{equation}

Consider all the tree topologies that can be assembled from $s$ nodes.
The weight of each tree topology, $\tau$,
is given by the number of ways to distribute $s$ indistinguishable nodes 
among its branches and arms.
We denote this weight by $W(\tau;s)$.
In the case of a tree that consists of a linear chain of $s$ nodes,
there are no degrees of freedom.
Therefore, its weight is

\begin{equation}
W[\tau=(0,\cdot,2);s]=1.
\end{equation}

\noindent
The weight of a tree of $s$ nodes
that consists of a single hub and $b$ branches
is given by

\begin{equation}
W[\tau=(1,0,b);s]=
\binom{s-2}{b-1}.
\end{equation}

\noindent
Here
the binomial factor 
counts the number of possibilities
to distribute 
$s-1$ nodes between the $b$ branches,
such that each branch
consists of at least one node. 
The weight of a tree with two hubs is

\begin{equation}
W[\tau=(2,A,\vec b);s]=  
\binom{s-2}{b}.
\label{eq:W}
\end{equation}

\noindent
In this case 
the binomial coefficient accounts for the number of ways to
distribute $s-2$ nodes between the $b$ branches and one arm,
where each branch includes at least one node.

In general, the weight
of a tree structure consisting of $s$ nodes, $h$ hubs 
(connected by $h-1$ arms) and a branch vector $\vec b$ is

\begin{equation}
W[\tau=(h,A,\vec b);s] = \binom{s-2}{h+b-2}.
\label{eq:Wtaus}
\end{equation}

\noindent
This result can be understood as follows.
Among the $s$ nodes, $h$ nodes are fixed as hubs while each one of the $b$ branches
includes at least one node. Therefore, there are $x=s-h-b$ nodes that can be
distributed among the $y=b+h-1$ branches and arms. 
Using Eq. (\ref{eq:partitions}) for the
number of possible
divisions of $x$ objects among $y$ boxes,
one obtains the result of  
Eq. (\ref{eq:Wtaus}).

The contribution of each tree topology 
to the statistical properties of the network such as the 
degree distribution and the DSPL
also depends on its symmetry.
To account for the effect of the symmetry, 
we define the symmetry factor

\begin{equation}
X(\tau)= \frac{1}{|{\rm Aut}(\tau)|},
\end{equation}

\noindent
where
${\rm Aut}(\tau)$ is
the automorphism group of $\tau$
\cite{Bollobas2001},
namely all the transformations that
leave $\tau$ unchanged.
It can be expressed as a product of the form

\begin{equation}
|{\rm Aut}(\tau)| = |{\rm Aut}(A)| |{\rm Aut}(\vec b)|,
\end{equation}

\noindent
where ${\rm Aut}(A)$ is the automorphism group of the 
backbone tree,
which consists of the hubs alone,
and ${\rm Aut}(\vec b)$ is the automorphism
group of the branches.
While $|{\rm Aut}(A)|$ depends on the overall symmetry of the
backbone tree, $|{\rm Aut}(\vec b)|$
is given by

\begin{equation}
|{\rm Aut}(\vec b)| = \prod_{i=1}^{h} b_i !.
\end{equation}

\noindent
For example, in the case of a linear chain of $s$ nodes,

\begin{equation}
X[\tau=(0,\cdot,2)]=\frac{1}{2}.
\end{equation}

\noindent
For a tree consisting of
a single hub of $b$ branches

\begin{equation}
X[\tau=(1,0,b)]=\frac{1}{b!},
\end{equation}

\noindent
while for a tree that includes two hubs
with $b_1$ and $b_2$ branches,

\begin{equation}
X[\tau=(2,A,\vec b)]= \frac{1}{2 b_1! b_2!}.
\end{equation}

\noindent
For a tree that consists of a central hub surrounded by
three peripheral hubs

\begin{equation}
X[\tau=(4,A,\vec b)]= \frac{1}{3 b_1! b_2! b_3! b_4!}.
\end{equation}

\section{The degree distribution}

In this Section we show how to use the topological expansion 
to express the degree distribution $P_{\rm FC}(K=k)$ as a composition of the
contributions of the different tree topologies.
In this case the asymptotic form is known to be the Poisson distribution,
$\pi(K=k)$, which enables us to validate the method.

Consider a tree that consists of $s \ge 2$ nodes,
whose degree sequence is given by $k_1,k_2,\dots,k_s$.
Since a tree of $s$ nodes includes $s-1$ edges,
the sum of these degrees satisfies

\begin{equation}
\sum_{i=1}^s k_i = 2(s-1).
\label{eq:sk_i}
\end{equation}

\noindent
We denote the number of nodes of degree $k$ by $N(K=k)$,
where

\begin{equation}
\sum_{k=1}^{s-1} N(K=k) = s,
\label{eq:sNk}
\end{equation}

\noindent
reflecting the fact that in a tree of size $s \ge 2$ the degrees
of all nodes satisfy $k \ge 1$.
In the special case of an isolated node,
$s=1$ and $N(K=k)=\delta_{k,0}$,
where $\delta_{k,k'}$ is the Kronecker delta,
which satisfies $\delta_{k,k'}=1$ if $k=k'$ and
$\delta_{k,k'}=0$ otherwise.

Eq. (\ref{eq:sk_i}) can be written in the form

\begin{equation}
\sum_{k=1}^{s-1} k N(K=k) = 2(s-1),
\label{eq:skNk}
\end{equation}

\noindent
where $s \ge 2$.
Combining 
Eqs. (\ref{eq:sNk}) and (\ref{eq:skNk}),
we obtain

\begin{equation}
N(K=1) = 2 + \sum_{k=3}^{s-1} (k-2) N(K=k).
\end{equation}

\noindent
This result reflects the fact that any tree includes at least two
leaf nodes and provides a relation between the degrees of the hubs
and the number of leaf nodes in a tree.
The number of nodes of degree $k=2$ can be obtained from

\begin{equation}
N(K=2) = s - N(K=1) - \sum_{k=3}^{s-1} N(K=k).
\end{equation}

\noindent
The topology of each tree structure can be 
described by
$\tau = (h,A,\vec b)$,
where 

\begin{equation}
h = \sum_{k=3}^{s-1} N(K=k)
\end{equation}

\noindent
is the number of hubs.
The degrees of the hubs are given by

\begin{eqnarray}
k_1 &=& a_1+b_1
\nonumber \\
k_2 &=& a_2+b_2
\nonumber \\
    &:& 
\nonumber \\
k_h &=& a_h+b_h,
\end{eqnarray}

\noindent
where $a_i$ is the number of arms and $b_i$ is the number of branches
that are connected to hub $i$.
The number of leaf nodes, with degree $k=1$ is given by $b=\sum_{i=1}^{h} b_i$.
The remaining $s-h-b$ nodes are of degree $k=2$,
namely

\begin{equation}
N(K=2) = s - h - b.
\end{equation}

The number of nodes of degree $k$ in a linear chain of $s$ nodes
is given by

\begin{equation}
N(K=k) = 2 \delta_{k,1} + (s-2) \delta_{k,2},
\end{equation}

\noindent
where $\delta_{k,k'}$ is the Kronecker delta.

The number of nodes of degree $k$ in a tree that consists of $s$ nodes, and includes
a single hub with $b$ branches, is

\begin{equation}
N(K=k) = b \delta_{k,1} + (s-1-b) \delta_{k,2} + \delta_{k,b}.
\end{equation}

\noindent
The number of nodes of degree $k$ in a tree that consists of
$s$ nodes, and takes the form of a chain of $h$ hubs, with a total of $b$
branches distributed according to $\vec b = (b_1,b_2,\dots,b_h)$, is

\begin{equation}
N(K=k) = b \delta_{k,1} + (s-h-b) \delta_{k,2} + \sum_{i=2}^{h-1} \delta_{k,b_i+2}
+ \delta_{k,b_1+1} + \delta_{k,b_h+1}.
\label{eq:P_k}
\end{equation}

Consider a tree of topology 
$\tau=(h,A,\vec b)$
that consists of $s$ nodes. 
Such tree includes $h$ hubs,
whose degrees in the backbone tree are given by
$\vec a = (a_1,a_2,\dots,a_h)$,
and their branch vector is
$\vec b = (b_1,b_2,\dots,b_h)$.
The number of nodes of degree $k$ is given by

\begin{equation}
N(K=k|\tau;s) = b \delta_{k,1} + (s-h-b) \delta_{k,2} 
+ \sum_{i=1}^{h} \delta_{k,a_i+b_i}.
\label{eq:Nk}
\end{equation}

\noindent
The degree distribution,
$P_{\rm FC}(K=k | \tau;S=s)$,
of trees of topology $\tau$, 
which consist of $s$ nodes,
is given by

\begin{equation}
P_{\rm FC}(K=k | \tau;S=s) =
\frac{N(K=k | \tau;s)}{  s  },
\label{eq:Pktaus}
\end{equation}

\noindent
where 
$N(K=k|\tau;s)$ is given by
Eq. (\ref{eq:Nk}).
In the analysis below
we use conditional degree distributions
that are evaluated under different conditions. 
In Table 
\ref{table:P}
we summarize these
distributions and list the equations from which they can be evaluated.

The degree distribution over all the tree topologies
that consist of $s$ nodes 
is given by

\begin{equation}
P_{\rm FC}(K=k | S=s) =
\frac{
\sum\limits_{ \{ \tau | s \}} X(\tau) W(\tau;s) P_{\rm FC}(K=k | \tau;S=s)
}{
\sum\limits_{ \{ \tau | s \}} X(\tau) W(\tau;s)
},
\label{eq:P_k2}
\end{equation}

\noindent
where $k=1,2,\dots,s-1$, the probabilities
$P_{\rm FC}(K=k | \tau;S=s)$
are given by 
Eq. (\ref{eq:Pktaus}) 
and the summation is over all component topologies that
can be constructed from $s$ nodes.

In Table 
\ref{table:Pkeqs}
we present the conditional degree distributions 
$P_{\rm FC}(K=k | S=s)$
for trees of $s=2,3,\dots,10$ nodes.
These distributions are
determined by the combinatorial considerations presented above, 
after identifying by hand all the tree topologies that 
appear in trees of size $2 \le s \le 10$. The probabilities are 
expressed in terms of constant rational numbers.

Summing up the degree distributions
obtained from 
Eq. (\ref{eq:P_k2})
over all the tree topologies that consist of $s' = 2,3,\dots,s$ nodes,
with suitable weights,
we obtain

\begin{equation}
P_{\rm FC}(K=k | 2 \le S \le s) =
\frac{ 
\sum\limits_{s'=2}^{s} 
s'
P_{\rm FC}(S=s') 
P_{\rm FC}(K=k | S=s')
}{
\sum\limits_{s'=2}^{s} 
s'
P_{\rm FC}(S=s')
}.
\label{eq:dspk3}
\end{equation}

\noindent
This equation provides an exact analytical expression for the degree distribution
over all tree topologies up to any desired size, $s$ (not including the case of
an isolated node).
In Table 
\ref{table:Pkles}
we present these expressions for 
$P_{\rm FC}(K=k |2 \le S \le s)$
where $s=2,3,\dots,6$
and 
$k=1,2,\dots,5$.
It turns out that in these expressions the dependence on the mean degree, $c$, always appears 
in terms of the parameter $\eta = \eta(c)$, which takes the form

\begin{equation}
\eta(c) = c e^{-c}.
\label{eq:eta}
\end{equation}

\noindent
The function $\eta(c)$ is a monotonically increasing function in the
interval $0 \le c \le 1$, where $\eta(0)=0$ and $\eta(1)=1/e$.
Expanding the results of
Eq. (\ref{eq:dspk3})
in powers of the small parameter $c$,
we obtain

\begin{equation}
P_{\rm FC}(K=k | 2 \le S \le s) 
= \frac{e^{-c} c^k}{(1 - e^{-c}) k!}  \left(1 + q_{s,k} c^{s-k} + \dots \right),
\label{eq:ddf}
\end{equation}

\noindent
where 
$k=1,2,\dots,s-1$ and the coefficients
$q_{s,k}$ are rational numbers of order $1$.

As $s$ is increased the degree distribution 
given by Eq. (\ref{eq:ddf})
converges to
the asymptotic form given by

\begin{equation}
\pi_{\rm FC}(K=k) = \frac{e^{-c} c^k}{(1 - e^{-c}) k!},
\label{eq:poisson_k>0}
\end{equation}

\noindent
which is the degree distribution
of the whole subcritical ER network,
except for the isolated nodes.
Taking into account the isolated nodes, whose weight in
the degree distribution is $\pi(K=0)=e^{-c}$, we obtain
the Poisson distribution introduced in Eq. (\ref{eq:poisson})

\begin{equation}
\pi(K=k) = e^{-c} \delta_{k,0} + (1-e^{-c}) \pi_{\rm FC}(K=k) \theta(k),
\end{equation}

\noindent
where $\theta(k)$ is the Heaviside function.

This convergence 
to the Poisson degree distribution
confirms the validity of the topological expansion
and shows that the combinatorial factors were evaluated correctly.
In Table 
\ref{table:Pkexp}
we present the leading correction terms,
$q_{s,k} c^{s-k}$, of Eq. (\ref{eq:ddf}),
obtained from the topological expansion, for all the
tree structures that consist of up to $s$ nodes, 
where $s=2,3,\dots,10$. 
Tree structures with up to $s$ nodes support degrees in the range of $k=1,\dots,s-1$.

\section{The mean and variance of the degree distribution}

The moments of the degree distribution provide useful information
about the network structure. The first and second moments are
of particular importance. The first moment, 
$\langle K \rangle_{\rm FC}$
provides the mean degree.
The width of the distribution is characterized by the
variance, 
${\rm Var}(K) = \langle K^2 \rangle - \langle K \rangle^2$,
where $\langle K^2 \rangle$ is the second moment.

The 
$n^{\rm th}$ moment of the degree distribution,
over all trees of topology $\tau$ that consist of $s$ nodes,
can be expressed by

\begin{equation}
{\mathbb E}[K^n | \tau; S=s] = \sum_{k=1}^{s-1} k^n P_{\rm FC}(K=k |\tau; S=s),
\label{eq:E[K|tau_s]}
\end{equation}

\noindent
where 
$P_{\rm FC}(K=k |\tau; S=s)$
is given by Eq. (\ref{eq:Pktaus}).
The 
$n^{\rm th}$ moment of the degree distribution
over all tree topologies that consist of $s$ nodes
is given by

\begin{equation}
{\mathbb E}[K^n | S=s] = 
\frac{
\sum\limits_{ \{ \tau | s \}} X(\tau) W(\tau;s) {\mathbb E}[K^n | \tau;S=s]
}{
\sum\limits_{ \{ \tau | s \}} X(\tau) W(\tau;s)
},
\label{eq:E[K|s]}
\end{equation}

\noindent
where
${\mathbb E}[K^n | \tau;S=s]$ 
is given by
Eq. (\ref{eq:E[K|tau_s]}).
For the special case of $n=1$, 
one obtains

\begin{equation}
{\mathbb E}[K | S=s] = 2 - \frac{2}{s}.
\label{eq:Eks}
\end{equation}

\noindent
This result represents a topological invariance and it applies to any
tree of $s$ nodes, regardless of its topology, $\tau$.
This is due to the fact that any tree of $s$ nodes includes
$s-1$ edges and each edge is shared by two nodes. 
The results for the first two moments,
${\mathbb E}[K|S=s]$
and
${\mathbb E}[K^2|S=s]$,
and for the variance
${\rm Var}[K|S=s] = {\mathbb E}[K^2|S=s] - ({\mathbb E}[K|S=s])^2$,
for $s=2,3,\dots,10$
are shown in Table
\ref{table:Pkeqs}.

The $n^{\rm th}$ moment of the degree
distribution over all trees that consist of up to $s$ nodes 
(except for the isolated nodes)
is given by

\begin{equation}
{\mathbb E}[K^n | 2 \le S \le s] 
= \frac{\sum\limits_{s'=2}^s s' P_{\rm FC}(S=s')  
{\mathbb E}[K^n | S=s']}{\sum\limits_{s'=2}^s s' P_{\rm FC}(S=s')},
\label{eq:E[K|Sles]}
\end{equation}

\noindent
where
$P_{\rm FC}(S=s')$
is given by Eq. (\ref{eq:Ps}).
Performing the summation for a given value of $s$ provides an exact analytical
expression for the $n^{\rm th}$ moment of the degree distribution
over all tree topologies that consist of up
to $s$ nodes.
The resulting expressions for the mean degree,
${\mathbb E}[K|2 \le S \le s]$,
over all trees that consist of up to $s=2,3,\dots,6$ nodes,
are presented in Table \ref{table:Pkles}.

For a tree of size $s=1$, which consists of a single, isolated node,
$P_{\rm FC}(S=1)=2 e^{-c}/(2-c)$
and ${\mathbb E}[K^n|S=1]=0$.
Thus in order to account for the isolated nodes, one should simply 
add the term $2 e^{-c}/(2-c)$ to the denominator of Eq. (\ref{eq:E[K|Sles]}).

In the limit of large $s$, the mean degree ${\mathbb E}[K |2 \le S \le s]$
converges towards the asymptotic result, which is given by

\begin{equation}
\langle K \rangle_{\rm FC} = \frac{c}{1 - e^{-c}}.
\label{eq:<K>}
\end{equation}

\noindent
Taking into account the isolated nodes, we obtain

\begin{equation}
\langle K \rangle = (1-e^{-c}) \langle K \rangle_{\rm FC} = c.
\label{eq:<K1>}
\end{equation}

\begin{figure}
\begin{center}
\includegraphics[width=12cm]{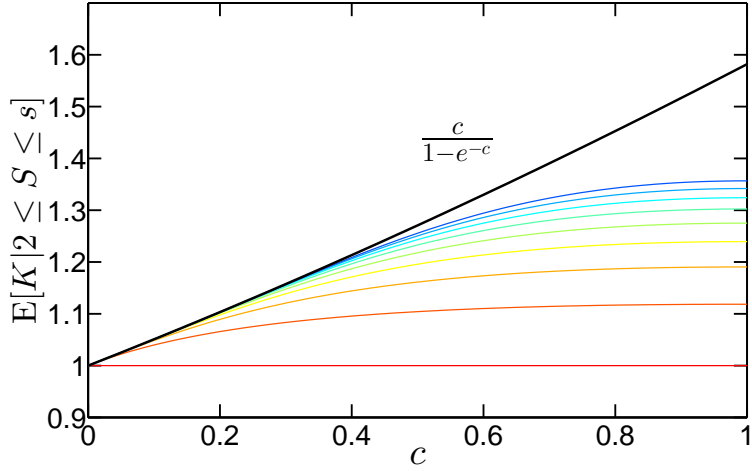}
\end{center}
\caption{
(Color online)
Analytical results for the
mean degree,
${\mathbb E}[K |2 \le S \le s]$,
over all tree topologies of sizes smaller
or equal to $s$, as a function of $c$,
for $s=2,3,\dots,10$
(solid lines), from bottom to top, respectively,
obtained from Eq. (\ref{eq:EK2Ss}).
The thick solid line shows the asymptotic result,
$\langle K \rangle_{\rm FC}$.
}
\label{fig:4}
\end{figure}

In Fig. \ref{fig:4} we present the
mean degrees,
${\mathbb E}[K | 2 \le S \le s]$,
as a function of $c$
(thin solid lines).
The results are shown for
all tree topologies of sizes smaller
or equal to $s$, 
where $s=2,3,\dots,10$
(from bottom to top).
The thick solid line shows the asymptotic result,
$\langle K \rangle_{\rm FC}$,
given by Eq. (\ref{eq:<K>}).

Below we derive closed form analytical expressions for the
mean degree, 
${\mathbb E}[K|2 \le S \le s]$,
over all tree topologies that consist of at least two nodes and up to $s$ nodes.
Trees of size $s=1$, which consist of isolated nodes, are excluded from this summation because 
the degree of such nodes is $k=0$, while trees of size $s \ge 2$ do not include nodes
of zero degree.
Inserting the expression for ${\mathbb E}[K|S=s]$
from Eq. (\ref{eq:Eks}) into equation (\ref{eq:E[K|Sles]}),
with $n=1$, we obtain

\begin{equation}
{\mathbb E}[K|2 \le S \le s] =
2 - 2 \frac{\sum\limits_{s'=2}^s P_{\rm FC}(S=s')}{\sum\limits_{s'=2}^s s' P_{\rm FC}(S=s')}.
\end{equation}

\noindent
This result can be expressed in the form

\begin{equation}
{\mathbb E}[K|2 \le S \le s] =
2 -  \frac{2 - c - 2 e^{-c} - 2 \sum\limits_{s'=s+1}^{\infty} P_{\rm FC}(S=s')}
{1 - e^{-c} - \sum\limits_{s'=s+1}^{\infty} s' P_{\rm FC}(S=s')}.
\label{eq:Eksum}
\end{equation}

\noindent
Expressing the distribution
$P_{\rm FC}(S=s')$
by Eq. (\ref{eq:Ps}) we obtain

\begin{equation}
{\mathbb E}[K|2 \le S \le s] =
2 -  \frac{ \sqrt{2 \pi} ( 2 - c -  2 e^{-c} )
- c^s e^{-(c-1)(s+1)} \Phi \left( c e^{1-c}, \frac{5}{2}, s+1 \right) }
{\sqrt{2 \pi} (1 - e^{-c}) - c^s e^{-(c-1)(s+1)} \Phi \left( c e^{1-c}, \frac{3}{2}, s+1 \right)},
\label{eq:Ekphi}
\end{equation}

\noindent
where $\Phi(z,s,a)$ is the Hurwitz Lerch $\Phi$ transcendent.
An alternative approach for the evaluation of 
${\mathbb E}[K|2 \le S \le s]$
is to go back to Eq. (\ref{eq:Eksum})
and replace the sums, $\sum\limits_{s'=s+1}^{\infty}$ by integrals of the form
$\int\limits_{s+1/2}^{\infty}$.
Performing the integrations, we obtain

\begin{equation}
{\mathbb E}[K|2 \le S \le s] =
2 -  \frac{ \sqrt{2 \pi} c( 2 - c - 2 e^{-c} )
- {(c-1-\ln c)^{3/2}}
\Gamma \left[ - \frac{3}{2},(c-1-\ln c)\left(s + \frac{1}{2} \right) \right] }
{\sqrt{2 \pi} c (1 - e^{-c}) - (c-1-\ln c)^{1/2}
\Gamma \left[ - \frac{1}{2},(c-1-\ln c)\left(s + \frac{1}{2} \right) \right]},
\label{eq:EK2Ss}
\end{equation}

\noindent
where $\Gamma(s,a)$ is the incomplete Gamma function.
This function satisfies

\begin{equation}
\Gamma \left( - \frac{3}{2},x \right) =
\frac{4}{3} \sqrt{\pi} \left[1 - {\rm erf}(\sqrt{x}) \right]
+ \frac{2 e^{-x} (1-2x)}{3 x^{3/2}},
\end{equation}

\noindent
and

\begin{equation}
\Gamma \left( - \frac{1}{2},x \right) =
- 2 \sqrt{\pi} \left[1 - {\rm erf}(\sqrt{x}) \right]
+ \frac{2 e^{-x}}{\sqrt{x}},
\end{equation}

\noindent 
where ${\rm erf}(x)$ is the error function.
In the limit of $c \rightarrow 0$ one can show that
${\mathbb E}[K|2 \le S \le s] \rightarrow 1$.

For $c=1$ the $\Phi$ transcendent function in Eq. (\ref{eq:Ekphi})
can be replaced by the Hurwitz Zeta function. 
In this case

\begin{equation}
{\mathbb E}[K|2 \le S \le s] =
2 
-  \frac{ \sqrt{2 \pi} ( 2 e - 4  )
- 4 e
\zeta \left( \frac{5}{2},s+1 \right)  }
{ \sqrt{2 \pi} (e - 1 )
- e
\zeta \left( \frac{3}{2},s+1 \right) },
\label{eq:Ekzeta}
\end{equation}

\noindent
where $\zeta(s,a)$ is the Hurwitz zeta function.
In the limit of large $s$, one can approximate Eq. (\ref{eq:Ekzeta})
by an asymptotic expansion of the form

\begin{equation}
{\mathbb E}[K|2 \le S \le s] =
\frac{e}{e-1} 
- \sqrt{ \frac{2}{\pi} }
\frac{e(e-2)}{(e-1)^2}
\frac{1}{\sqrt{s}}
-
\frac{2}{\pi} \frac{e^2(e-2)}{(e-1)^3} 
\frac{1}{s}
+ {\mathcal O} \left( \frac{1}{s^{3/2}} \right).
\label{eq:Ekzetaexp}
\end{equation}

\begin{figure}
\begin{center}
\includegraphics[width=12cm]{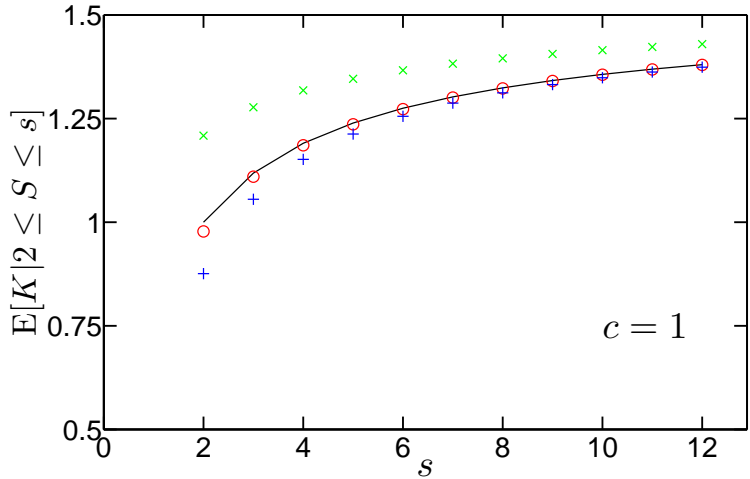}
\end{center}
\caption{
(Color online)
The mean degree 
${\mathbb E}[K|2 \le S \le s]$
over all trees of size smaller or equal to $s$,
as a function of $s$ for $c=1$.
The analytical results (circles),
obtained from Eq. (\ref{eq:Ekzeta}),
are in excellent agreement with the exact results
of the asymptotic expansion (solid line).
The results of the asymptotic expansion to order $1/\sqrt{s}$
($\times$), obtained from the first two terms of 
Eq. (\ref{eq:Ekzetaexp}),
exhibit large deviations from the exact results, particularly for
small values of $s$.
However, an expansion to order $1/s$ obtained by including the third term in 
Eq. (\ref{eq:Ekzetaexp}),
greatly improves the results ($+$).
}
\label{fig:5}
\end{figure}

In Fig. \ref{fig:5} we present the
mean degree 
${\mathbb E}[K|2 \le S \le s]$
over all trees of $S \le s$ nodes,
as a function of $s$ 
at the critical value of
$c=1$.
The analytical results (circles),
obtained from Eq. (\ref{eq:Ekzeta}),
are in excellent agreement with the exact results
of the asymptotic expansion (solid line).
The only slight deviations are for $s=2$ and $3$, and they reflect the
fact that
Eq. (\ref{eq:Ekzeta})
is based on the Stirling expansion,
which becomes 
accurate for $s \ge 4$.
The results of the asymptotic expansion to order $1/\sqrt{s}$
($\times$), obtained from the first two terms of 
Eq. (\ref{eq:Ekzetaexp}),
exhibit large deviations from the exact results, particularly for
small values of $s$.
This means that next order correction should be taken into account,
at least for such small values of $s$.
Indeed, an expansion to order $1/s$ obtained by including the third term in 
Eq. (\ref{eq:Ekzetaexp}),
greatly improves the results ($+$).

Using Eq. (\ref{eq:E[K|Sles]})
one can obtain exact analytical expressions for the second moment
of the degree distribution over all trees of size $2 \le S \le s$. The results for small
trees, which consist of up to $s=2,3,\dots,6$ nodes,
are shown in Table \ref{table:Pkles}.
In the limit of large $s$, the second moment 
${\mathbb E}[K^2 | 2 \le S \le s]$
converges towards the asymptotic result, which is given by

\begin{equation}
\langle K^2 \rangle_{\rm FC} = \frac{c(c+1)}{1 - e^{-c}}.
\label{eq:<K^2a>}
\end{equation}

\noindent
Taking into account the isolated nodes, we obtain

\begin{equation}
\langle K^2 \rangle = (1-e^{-c}) \langle K^2 \rangle_{\rm FC} = c(c+1).
\label{eq:<K^2b>}
\end{equation}

The variance of the degree distribution over all trees that consist of up
to $s$ nodes is given by

\begin{equation}
{\rm Var}[K | 2 \le S \le s]= {\mathbb E}[K^2 | 2 \le S \le s] - ({\mathbb E}[K | 2 \le S \le s])^2.
\end{equation}

\noindent
Using the results presented 
in Table \ref{table:Pkles}
for the first and second moments of 
the degree distributions over small trees
of sizes $s=2,3,\dots,5$,
we obtain

\begin{eqnarray}
{\rm Var}[K | 2 \le S \le 2] &=&
0
\nonumber \\
{\rm Var}[K | 2 \le S \le 3] &=&
\frac{2 \eta +2 \eta^2}{(2 + 3 \eta )^2}
\nonumber \\
{\rm Var}[K | 2 \le S \le 4] &=&
\frac{ 18 \eta +78 \eta^2+  90 \eta^3  +   96 \eta^4   }
{\left( 6 + 9 \eta + 16 \eta^2  \right)^2}
\nonumber \\
{\rm Var}[K | 2 \le S \le 5] &=&
\frac{  288 \eta +1248 \eta^2 +  3960 \eta^3  +5016 \eta^4+ 6920 \eta^5   +  7500 \eta^6 }
{\left( 24  +36 \eta+  +64 \eta^2 +   125 \eta^3 \right)^2}.
\end{eqnarray}

In the limit of large $s$, the variance ${\rm Var}[K | 2 \le S \le s]$
converges towards the asymptotic result, 
$\sigma^2_{K, {\rm FC}} = {\rm Var}(K)$,
where $\sigma_{k, {\rm FC}}$ is the standard deviation of the
degree distribution over all the finite components.
The asymptotic variance is given by

\begin{equation}
\sigma^2_{K, {\rm FC}} =
{\rm Var}(K) = 
\frac{c}{1 - e^{-c}} - \frac{c^2 e^{-c}}{(1-e^{-c})^2}.
\label{eq:VarK}
\end{equation}

\noindent
Taking into account the isolated nodes,
we obtain

\begin{equation}
\sigma^2_{K} = \langle K^2 \rangle - (\langle K \rangle)^2 = c.
\end{equation}

\section{The distribution of shortest path lengths}

In this Section we apply the topological expansion to obtain the DSPL
of subcritical ER networks and to express it in terms of the contributions
of the different tree topologies.
Summing up the contributions for all possible tree topologies supported
by up to $s$ nodes, we express the DSPL as a power series in $c$,
and find its asymptotic form in the limit of $N \rightarrow \infty$.

For each value of $s=2,3,\dots$, we identify all the tree topologies, $\tau$,
supported by $s$ nodes.
For each one of these tree topologies,
and for 
$\ell=1,2,\dots,s-1$, 
we calculate the number, $N(L=\ell | \tau;s)$ 
of pairs of nodes that reside at a distance 
$\ell$ from each other.
We then sum up these contributions over all the possible ways
to assemble $s$ nodes into the given tree topology. 
Below we describe the enumeration of the shortest paths
for a few simple examples of tree topologies.

In a linear chain of $s$ nodes there are $s-\ell$ pairs of 
nodes at distance $\ell$ from each other.
Therefore,

\begin{equation}
N[L=\ell | \tau=(0,\cdot,2);s] = 
\binom{s-\ell}{1}
\end{equation}

\begin{figure}
\begin{center}
\includegraphics[width=8.5cm]{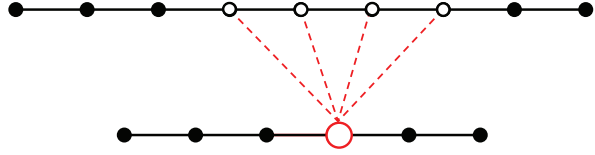}
\caption{
(Color online)
Illustration of the collapse process that
is used in order to obtain the combinatorial
factors for the DSPL on a finite component. 
In this case, the number of pairs of nodes that are
at a distance of $\ell=3$ from each other on a 
linear chain of size $s=9$ (top chain) is equal to the number of possible
locations of the marked node (large empty circle)
on the reduced chain of $s-\ell=6$ nodes
(bottom chain).
}
\label{fig:6}
\end{center}
\end{figure}

\noindent
A convenient way to evaluate the number of such 
pairs is to take a pair of nodes at
a distance $\ell$ from each other 
and reduce the chain of $\ell+1$ nodes between them
into a single node, which is marked in order to
distinguish it from the other nodes. 
This results in a reduced network of $k-\ell$ nodes,
one of which is the marked node. 
At this point, counting the number of pairs of nodes that 
are at a distance $\ell$ from each other is equivalent to counting the number
of different locations of the marked node in the
reduced network.
In Fig. \ref{fig:6} we illustrate this procedure for the case
of a linear chain of nodes.
Since each node in the reduced chain may be the marked node,
one concludes that in the original chain
there are $s-\ell$ pairs of nodes at a distance $\ell$ from each other.

For a tree of $s$ nodes that includes a single hub and $b$ 
branches, the number of pairs of nodes at a distance $\ell$ from
each other is

\begin{equation}
N[L=\ell | \tau=(1,0,b),s] = 
b \binom{s-\ell}{b}
+ (\ell-1) 
\binom{b}{2}
\binom{s-\ell}{b-1}.
\label{eq:dspl1b}
\end{equation}

\noindent
In this case
there are many different configurations 
due to the different ways to distribute
the nodes between the $b$ branches. 
Therefore, we need to sum up the numbers of
pairs of nodes at distance $L=\ell$ 
from each other in all the different configurations.
In this calculation we distinguish between pairs of nodes
that reside on the same branch
and pairs of nodes that reside on 
different branches.
To calculate the number of pairs of nodes 
residing on the same branch and are at a distance
$L=\ell$ apart, we pick one such pair of nodes and reduce the 
chain of $\ell+1$ nodes between them into
a single node. 
This node is marked in order to keep track of its location. 
The reduced network now consists of $s-\ell$ nodes. 
We then evaluate the number
of ways to distribute these $s-\ell$ 
nodes between the $b$ branches 
and the number of ways to place
the marked node in its own branch. 
Essentially, the marked node splits its branch into two parts.
This means that the number of possible 
configurations is equal to the number of possible ways
to distribute $s-\ell$ nodes to $b+1$ urns.
The first binomial coefficient in Eq.
(\ref{eq:dspl1b}) 
accounts for the number of 
such distributions.

To calculate the number of pairs of nodes 
that reside on different branches and
are at a distance $\ell$ apart from each other, 
we first arrange all $s$ nodes in a linear
chain. We choose a pair of nodes that are at a distance $\ell$ 
from each other and
reduce the $\ell+1$ nodes between them into a single node. 
This results in a reduced chain
of $s-\ell$ nodes, one of which is the marked node. 
We proceed in two stages. In the first
stage we consider the two branches on which the nodes $i$ and $j$ 
reside as a single branch,
which now includes the marked node. 
The binomial coefficient 
$\binom{s-\ell}{b-1}$
accounts for
the number of ways to distribute the nodes 
into $b-1$ urns and to choose randomly
the location of the marked node. 
In the second stage we randomly choose the location
of the hub among the $\ell-1$ 
nodes between $i$ and $j$ and connect all the end-points
of all other 
$b-2$ branches to this node. 
Apart from
this, there are 
$\binom{b}{2}$ 
possible ways to choose the branches on which $i$ and $j$ are
located.

The approach presented above can also be used to evaluate the number of pairs of nodes
at a distance $L=\ell$ apart that reside on branches that do not share a hub.
In this case one needs to account for the number of possible ways to locate two or
more hubs along the segment of $\ell-1$ nodes between $i$ and $j$.
For a tree of $s$ nodes, which includes two hubs, we obtain

\begin{eqnarray}
N[L&=&\ell | \tau=(2,A,\vec b);s] = 
(b_1+b_2+1) \binom{s-\ell}{b_1+b_2+1}
\nonumber \\
&+& (\ell-1) \left[ { \binom{b_1+1}{2} } + { \binom{b_2 +1}{2} } \right]
{ \binom{s-\ell}{b_1+b_2} }    
\nonumber \\
&+&b_1 b_2 { \binom{\ell-1}{2} }  { \binom{s-\ell}{b_1+b_2 - 1} }, 
\label{eq:dspl1}
\end{eqnarray}

\noindent
where $A$ is given by Eq. (\ref{eq:2by2})
and $\vec b = (b_1,b_2)$.
Generalizing this result to the case of a linear chain of $h$ hubs
we obtain

\begin{eqnarray}
N[L&=&\ell | \tau=(h,A,\vec b),s] =
(b+h-1) \binom{\ell-1}{0} \binom{s-\ell}{b+h-1}
\nonumber \\
&+&
\left[ \binom{b_1+1}{2} + \sum_{i=2}^{h-1} \binom{b_i+2}{2} + \binom{b_h+1}{2} \right]
\binom{\ell-1}{1} \binom{k-\ell}{b+h-2}
\nonumber \\
&+&
\sum_{r=2}^{h-1}  \left[ b_1(b_{r+1}+1) 
+ 
\sum_{i=1}^{h-r-1} (b_i+1)(b_{i+r}+1) 
+ 
(b_{h-r}+1)b_h \right] 
\nonumber \\
&\times&
\binom{\ell-1}{r} \binom{s-\ell}{b+h-r-1}
+
b_1 b_h \binom{\ell-1}{h} \binom{s-\ell}{b-1},
\end{eqnarray}

\noindent
where $A$ is an $h \times h$ Toeplitz matrix that satisfies
$A_{ij} = 1$ if $j = i \pm 1$ and $A_{ij}=0$ otherwise.
Similarly, for a tree that consists of a central hub, which is
surrounded by $h-1$ peripheral hubs, 
$N(L=\ell | \tau;s)$ is given by

\begin{eqnarray}
N[L&=&\ell | \tau=(h,A,\vec b);s] =
(b+h-1) \binom{\ell-1}{0} \binom{s-\ell}{b+h-1}
\nonumber \\
&+&
\left[ \binom{b_1+h-1}{2} + \sum_{i=2}^{h} \binom{b_i+1}{2}  \right]
\binom{\ell-1}{1} \binom{s-\ell}{b+h-2}
\nonumber \\
&+&
(b_1+h-2) \sum_{i=2}^{h} b_i
\binom{\ell-1}{2} \binom{s-\ell}{b+h-3}
\nonumber \\
&+&
\sum_{i=2}^{h} \sum_{j=i+1}^{h}
b_i b_j \binom{\ell-1}{3} \binom{s-\ell}{b+h-4},
\end{eqnarray}

\noindent
where $A_{1j}=1$ for $j \ge 2$,
$A_{i1}=1$ for $i \ge 2$
and $A_{ij}=0$ otherwise.

We will now derive an equation for the number of pairs of nodes
at a distance $\ell$ from each other in any given tree of $s$ nodes,
whose structure is given by the topology $\tau=(h,A,\vec b)$.
Such tree includes $h$ hubs, whose degrees are given by
the vector 

\begin{equation}
\vec k = (k_1,k_2,\dots,k_h), 
\end{equation}

\noindent
where
$k_i = a_i + b_i$.
For convenience we also define the vector

\begin{equation}
\vec k' = (k_1-1,k_2-1,\dots,k_h-1).
\end{equation}

\noindent
The hubs form a backbone tree of $h$ nodes,
described by the adjacency matrix, $A$, of dimensions $h \times h$.
For any pair of hubs, $i$ and $j$,
which are connected by an arm
(regardless of its length in the complete tree), 
the matrix element
$A_{ij}=1$, while otherwise $A_{ij}=0$.
From the adjacency matrix, $A$, one can obtain the $h \times h$ distance matrix,
$D$, of the backbone tree, which consists of the hubs alone. 
This is a symmetric matrix, whose matrix element $D_{ij}$ is the distance between
hub $i$ and hub $j$ on the backbone tree,
and the diagonal elements are $D_{ii}=0$.
For the analysis presented below, it is useful to express the distance matrix
as a sum of symmetric binary matrices in the form

\begin{equation}
D = D_{\bf 1} + 2 D_{\bf 2} + 3 D_{\bf 3} + \dots + (h-1) D_{h-1},
\end{equation}

\noindent
where
$(D_{\ell})_{ij}=1$ if 
$D_{ij}=\ell$ and 
$(D_{\ell})_{ij}=0$ otherwise.
The matrix $D_{\ell}$,
$\ell=1,2,\dots,h-1$
is called the $\ell^{\rm th}$ order vertex-adjacency matrix
\cite{Janezic2015}.
It can be obtained directly from the adjacency 
matrix, $A$, by constructing its powers $A^1$, $A^2$, $\dots$, $A^{\ell}$.
In case that
$(A^{\ell})_{ij} \ge 1$,
under the condition that
$(A^{\ell'})_{ij}=0$ 
for all the lower powers of $A$, namely
$\ell'=1,2,\dots,\ell-1$,
then
$(D_{\ell})_{ij}=1$,
and otherwise
$(D_{\ell})_{ij}=0$.

Each pair of nodes, $i$ and $j$ in the network can be classified according
to the number of hubs, $\nu_{ij}$, along the path between them.
If $i$ and $j$ reside on the same branch or on the same arm, $\nu_{ij}=0$.
If they reside on different branches or arms that emanate from the same
hub, $\nu_{ij}=1$. 
In case that $i$ and $j$ reside on branches or arms that do not
share a hub, 
we denote by $h_i$ the hub that is nearest to $i$ along the
path to $j$ and by $h_j$ the hub that is nearest to $j$ along the path to $i$.
We denote by $D_{ij}$ the distance between the hubs $h_i$ and $h_j$
on the backbone tree, which consists of the hubs alone.
The number of hubs along the shortest path between nodes $i$ and 
$j$ can be expressed by $\nu_{ij} =D_{ij}+1$.
Thus, $\nu_{ij}$ may take values in the range $0 \le \nu_{ij} \le h$.

The number of pairs of nodes that are at a distance $\ell$ from 
each other can be expressed in the form

\begin{equation}
N(L=\ell | \tau,s) =
\sum_{\nu=0}^{h} N_{\nu}(L=\ell | \tau,s),
\end{equation}

\noindent
where
$N_{\nu}(L=\ell | \tau,s)$
is the number of pairs of nodes, $i$ and $j$ that
are at a distance $\ell$ from each other,
and along the
path between them there are $\nu$ hubs.

For pairs of nodes that reside on the same branch or on
the same arm, for which $\nu=0$, we obtain

\begin{equation}
N_0(L=\ell | \tau,s) =
(b+h+1) \binom{\ell-1}{0} \binom{s-\ell}{b+h-1}.
\end{equation}

\noindent
For pairs of nodes that reside on different branches or arms
that emanate from the same hub, for which
$\nu=1$, we obtain

\begin{equation}
N_1(L=\ell | \tau,s) =
\binom{\ell-1}{1} \binom{s-\ell}{b+h-2}
\sum_{i=1}^h \binom{k_i}{2}.
\end{equation}

\noindent
For pairs of nodes for which $\nu \ge 2$ we obtain

\begin{equation}
N_{\nu}(L=\ell | \tau,s) =
\frac{1}{2}
\binom{\ell-1}{\nu} \binom{s-\ell}{b+h-\nu-1}
\sum_{i=1}^{h} \sum_{j=1}^{h}
k'_i k'_j \delta_{D_{ij},\nu-1}.
\label{eq:N_H}
\end{equation}

\noindent
Eq. (\ref{eq:N_H}) 
can be written in the form

\begin{equation}
N_{\nu}(L=\ell | \tau,s) =
\frac{1}{2}
\binom{\ell-1}{\nu} \binom{s-\ell}{b+h-\nu-1}
\vec k'^{T} D_{\nu-1} \vec k',
\label{eq:N_H2}
\end{equation}

\noindent
where $\vec k'^{T}$ is the transpose of $\vec k'$.

The distribution
$P_{\rm FC}(L = \ell | \tau; L < \infty,S=s)$,
for trees of a given topology, $\tau$, 
assembled from $s$ nodes,
is given by

\begin{equation}
P_{\rm FC}(L = \ell | \tau;L<\infty,S=s) =
\frac{N(L=\ell | \tau;s)}{ { \binom{s}{2}} W(\tau;s) }.
\label{eq:dspltaus}
\end{equation}

\noindent
In the analysis below
we use different types of DSPLs. 
In Table 
\ref{table:P}
we summarize these
distributions and list the equations from which each 
one of them can be evaluated.

The DSPL over components of all topologies
that consist of $s$ nodes, 
is given by

\begin{equation}
P_{\rm FC}(L=\ell |L<\infty, S=s) =
\frac{
\sum\limits_{ \{ \tau | s \}} X(\tau) W(\tau;s) P_{\rm FC}(L=\ell | \tau; L < \infty, S=s)
}{
\sum\limits_{ \{ \tau | s \}} X(\tau) W(\tau;s)
},
\label{eq:dspl2}
\end{equation}

\noindent
where the summation is over all component topologies which
can be constructed from $s$ nodes,
In Table 
\ref{table:Pleqs}
we present the probabilities
$P_{\rm FC}(L=\ell | L<\infty,S=s)$
for trees of $s=2,3,\dots,10$ nodes.
These probabilities 
are determined by combinatorial considerations and
are expressed in terms of
constant rational numbers.

To obtain the DSPL
over all the components of sizes 
$s' \le s$,
we sum up
the results of Eq.
(\ref{eq:dspl2})
over all these components: 

\begin{equation}
P_{\rm FC}(L=\ell | L<\infty, S \le s) =
\frac{ 
\sum\limits_{s'=2}^{s} 
\binom{s'}{2}
P_{\rm FC}(S=s')
P_{\rm FC}(L=\ell | L<\infty , S=s')
}{
\sum\limits_{s'=2}^{s} 
\binom{s'}{2}
P_{\rm FC}(S=s')
}.
\label{eq:dspl3}
\end{equation}

\noindent
This equation provides an exact analytical expression for the degree distribution
over all tree topologies up to any desired size, $s$.
In Table 
\ref{table:Plles}
we present these expressions for 
$P_{\rm FC}(L=\ell | L < \infty, S \le s)$
where $s=2,3,\dots,6$
and 
$\ell=1,2,\dots,5$.
It turns out that in these expressions the mean degree, $c$, always appears 
in the form $\eta (c)= c e^{-c}$, which is defined in Eq. (\ref{eq:eta}).

Expanding the results of 
Eq. (\ref{eq:dspl3})
as a power series in the small parameter $c$,
we find that

\begin{equation}
P_{\rm FC}(L=\ell | L<\infty, S \le s) = (1-c)c^{\ell-1} \left( 1 + r_{s,\ell} c^{s-\ell} + \dots \right), 
\label{eq:r_sell}
\end{equation}

\noindent
where $\ell=2,3,\dots,s-1$
and the coefficient
$r_{s,\ell}$ is a rational number of order $1$.
In Table 
\ref{table:Plexp}
we present the leading finite size correction terms,
$r_{s,\ell} c^{s-\ell}$, of Eq. (\ref{eq:r_sell}),
obtained from the topological expansion, for all the
tree structures that consist of up to $s$ nodes, 
where $s=2,3, \dots,10$. 
Tree structures with up to $s$ nodes support distances in the range of $\ell=1,\dots,s-1$. 
In the limit of large $s$, these
results converge towards the asymptotic form 

\begin{equation}
P_{\rm FC}(L = \ell | L<\infty) = (1-c)c^{\ell-1},
\label{eq:DSPL}
\end{equation}

\noindent
which turns out to be the DSPL of the entire subcritical network in the 
asymptotic limit of $N \rightarrow \infty$.
In spite of its apparent simplicity, this is a surprising and nontrivial result,
which was not anticipated when we embarked on the topological expansion.
Eq. (\ref{eq:DSPL}) is essentially a mean field result. 
Normally, a mean field result for the DSPL is expected to
represent the shell structure around a typical node.
However, in this case there is no typical node.
The shell structures around each node strongly depends on 
the size and topology of the component on which it resides as well
as on its location in the component.
Only by combining the contributions of all pairs of nodes one
obtains the simple expression of Eq. (\ref{eq:DSPL}).

The DSPL given by Eq. (\ref{eq:DSPL}) is a conditional distribution,
under the condition that the selected pair of nodes reside on the same component.
In fact, it is a subleading component of the overall DSPL of the network,
because in subcritical networks most pairs
of nodes reside on different components, and are thus at an infinite distance from each other.
The overall DSPL can be expressed by

\begin{equation}
P_{\rm FC}(L=\ell) = P_{\rm FC}(L < \infty) P_{\rm FC}(L=\ell | L<\infty),
\end{equation}

\noindent
where $P_{\rm FC}(L < \infty)$
is given by Eq. (\ref{eq:L<infty}).
Therefore,

\begin{equation}
P_{\rm FC}(L=\ell) = \frac{c^{\ell}}{N} 
\end{equation}

\noindent
for $\ell=1,2,\dots,N-1$, and

\begin{equation}
P_{\rm FC}(L = \infty) = 1 - \frac{c}{(1-c)N}.
\end{equation}

The tail distribution that corresponds to the probability distribution
function of Eq. (\ref{eq:DSPL}) is given by 

\begin{equation}
P_{\rm FC}(L>\ell | L<\infty) = c^{\ell}.
\label{eq:tail}
\end{equation}

\begin{figure}
\begin{center}
\includegraphics[width=12cm]{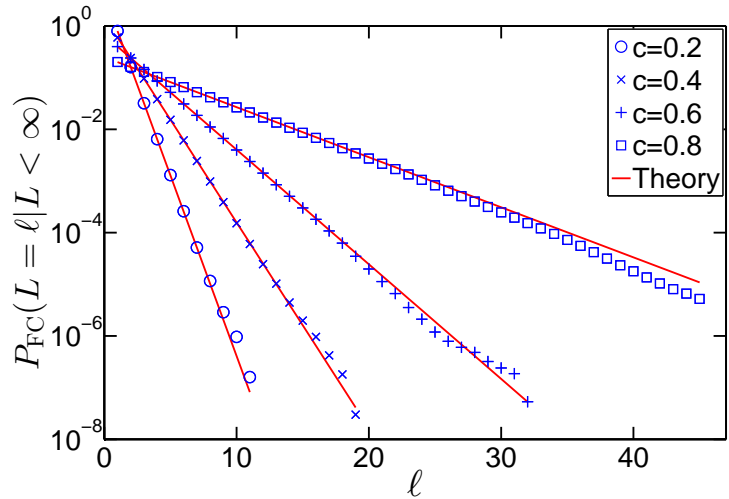}
\end{center}
\caption{
(Color online)
The DSPLs of subcritical ER network ensembles with 
$N=10^4$
and 
$c=0.2,0.4,0.6$ and $0.8$. 
The theoretical results for the corresponding asymptotic networks
(solid lines), obtained from
Eq.
(\ref{eq:DSPL})
are in very good agreement with the numerical simulations
(symbols).
The deviations in the tail are due to the
finite size of the sampled networks.
}
\label{fig:7}
\end{figure}

In Fig. 
\ref{fig:7}
we present theoretical results for the DSPL of asymptotic ER networks with
$c=0.2, 0.4, 0.6$ and $0.8$ (solid lines),
obtained from Eq. (\ref{eq:DSPL}).
These results are compared to 
numerical results for the DSPL (symbols), obtained
for networks of size $N=10^4$ and the
same four values of $c$. 
We find that the theoretical results are in very good
agreement with the numerical results except for
small deviations in the large distance tails.
These deviations are due to finite size of the simulated
networks.
The numerical simulations
were performed
via sampling of 
$10^4$ independent realizations of
ER networks of size $N=10^4$
for each value of $c$
\cite{Hartmann2015}.
For each realization we applied the 
\emph{all pairs shortest paths}
algorithm from the LEDA C++ library
\cite{Leda1999}.

\begin{figure}
\begin{center}
\includegraphics[width=8.0cm]{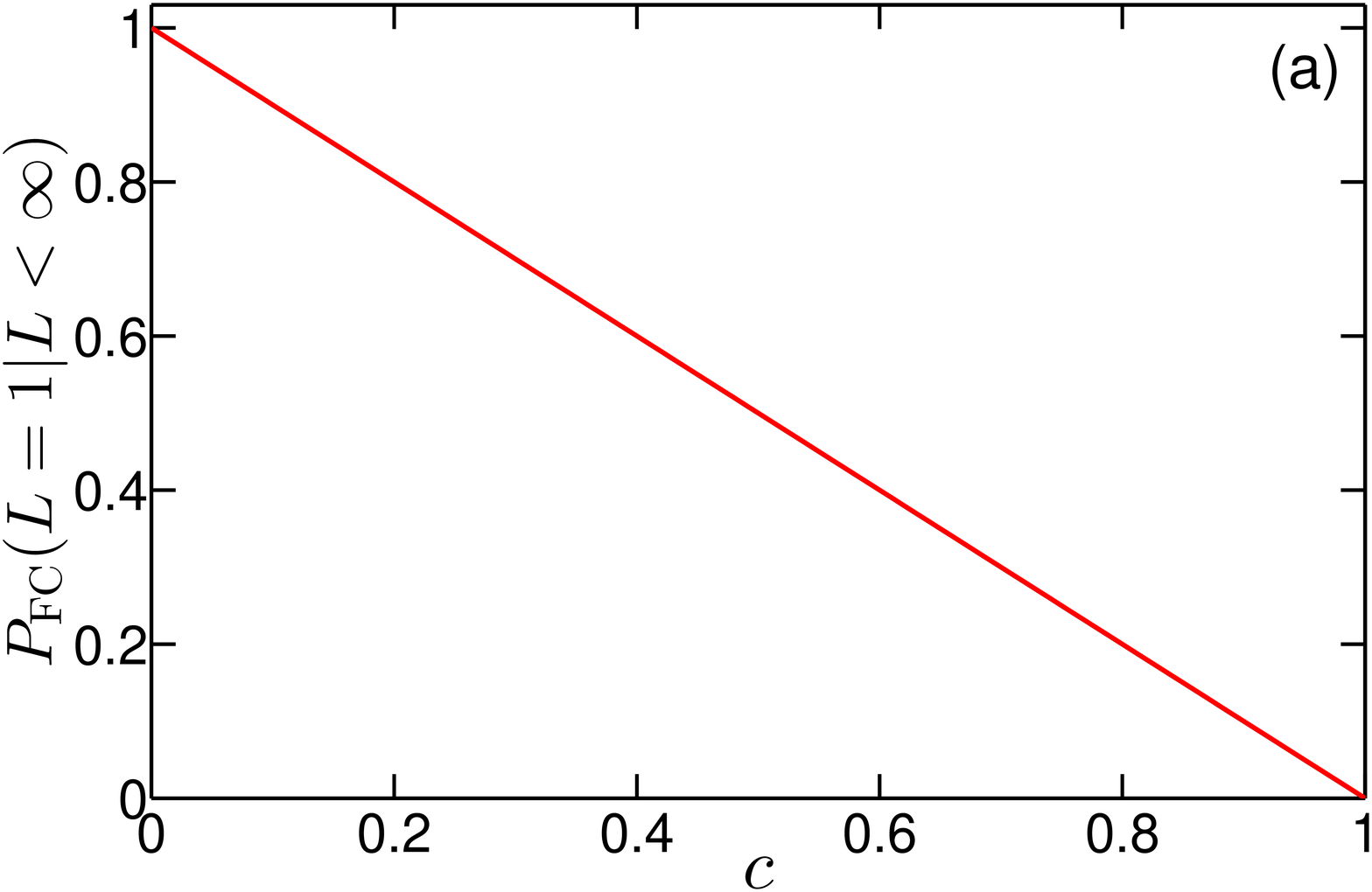} 
\includegraphics[width=8.0cm]{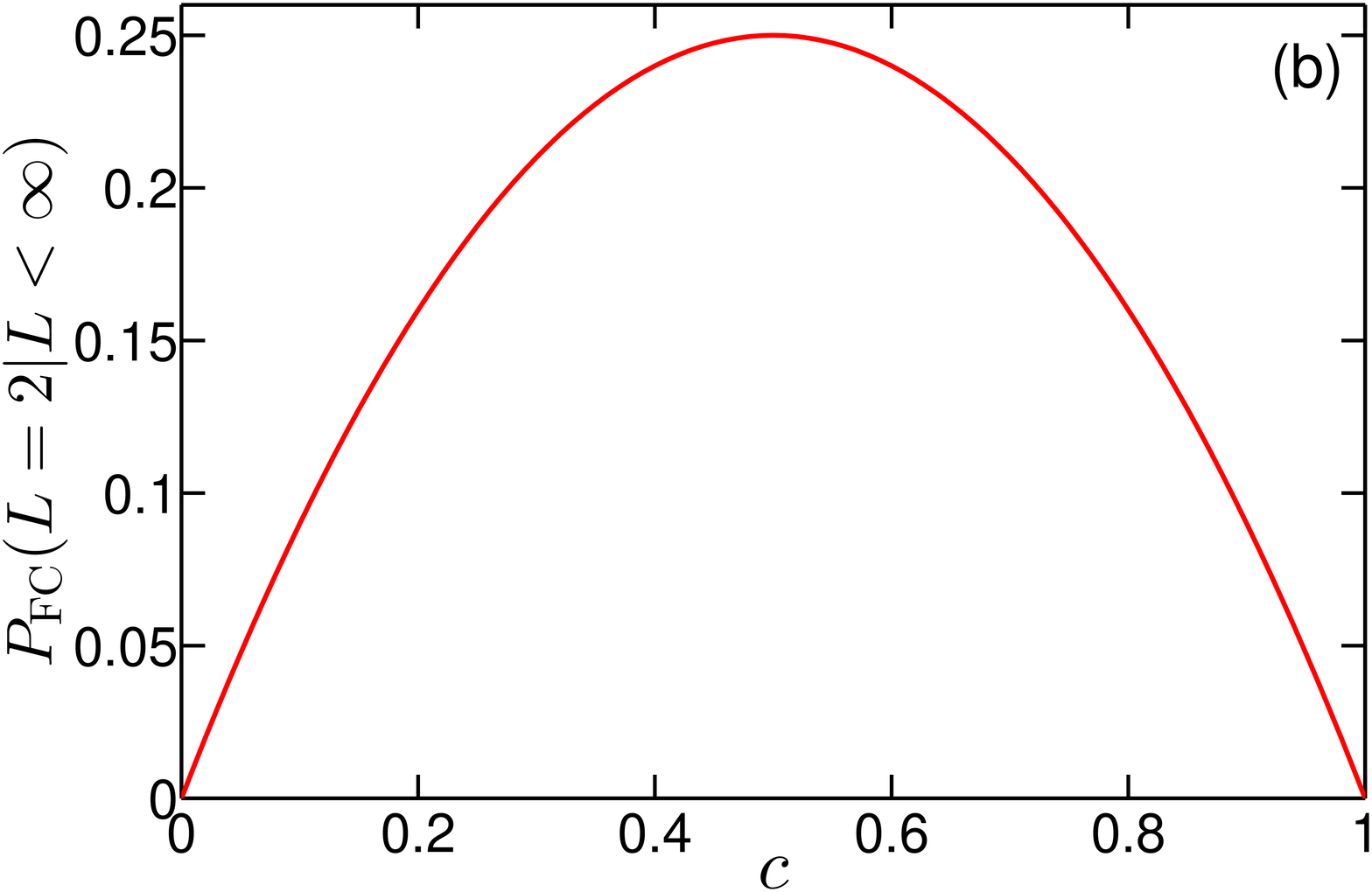} \\
\includegraphics[width=8.0cm]{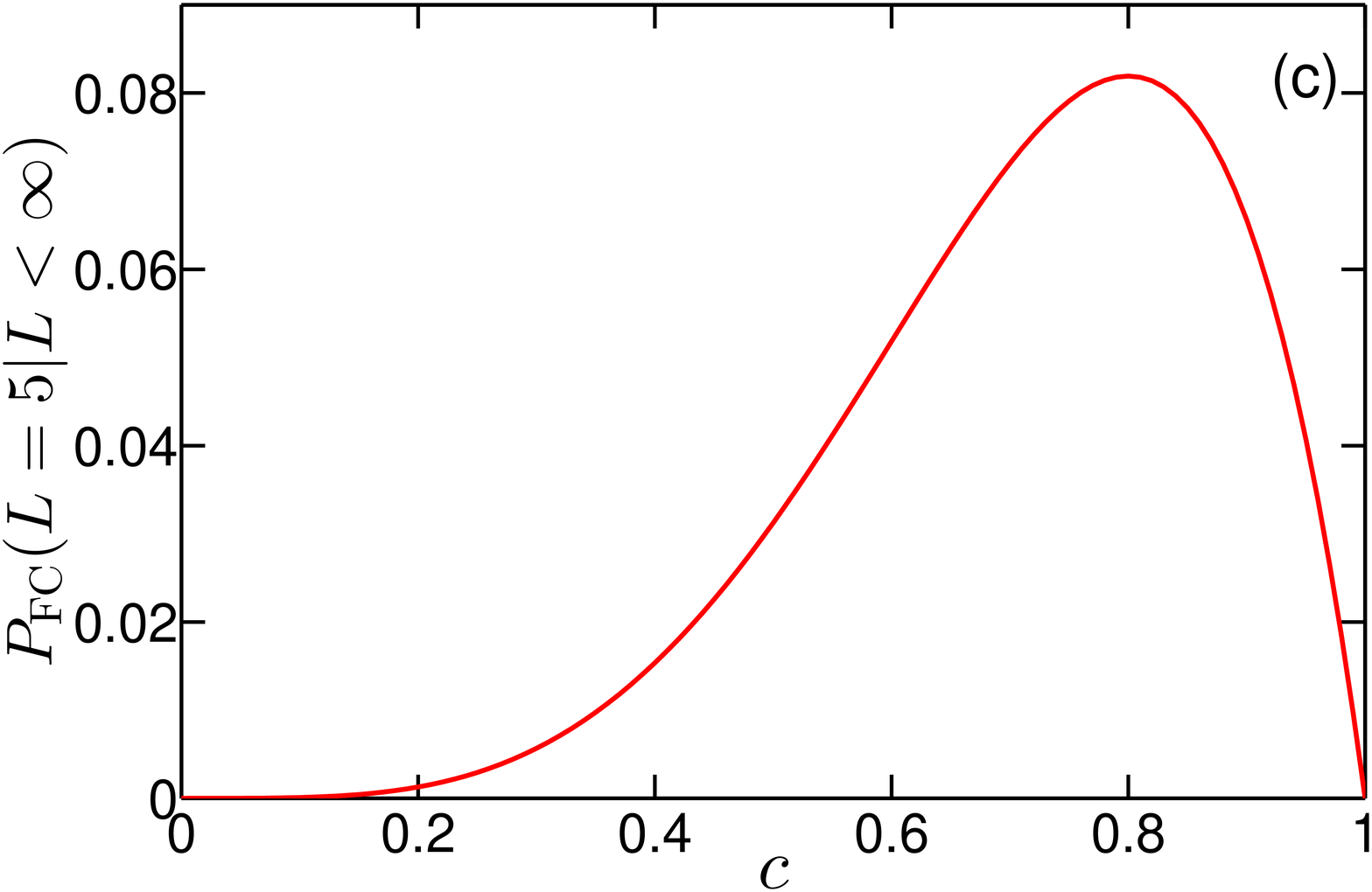} 
\includegraphics[width=8.0cm]{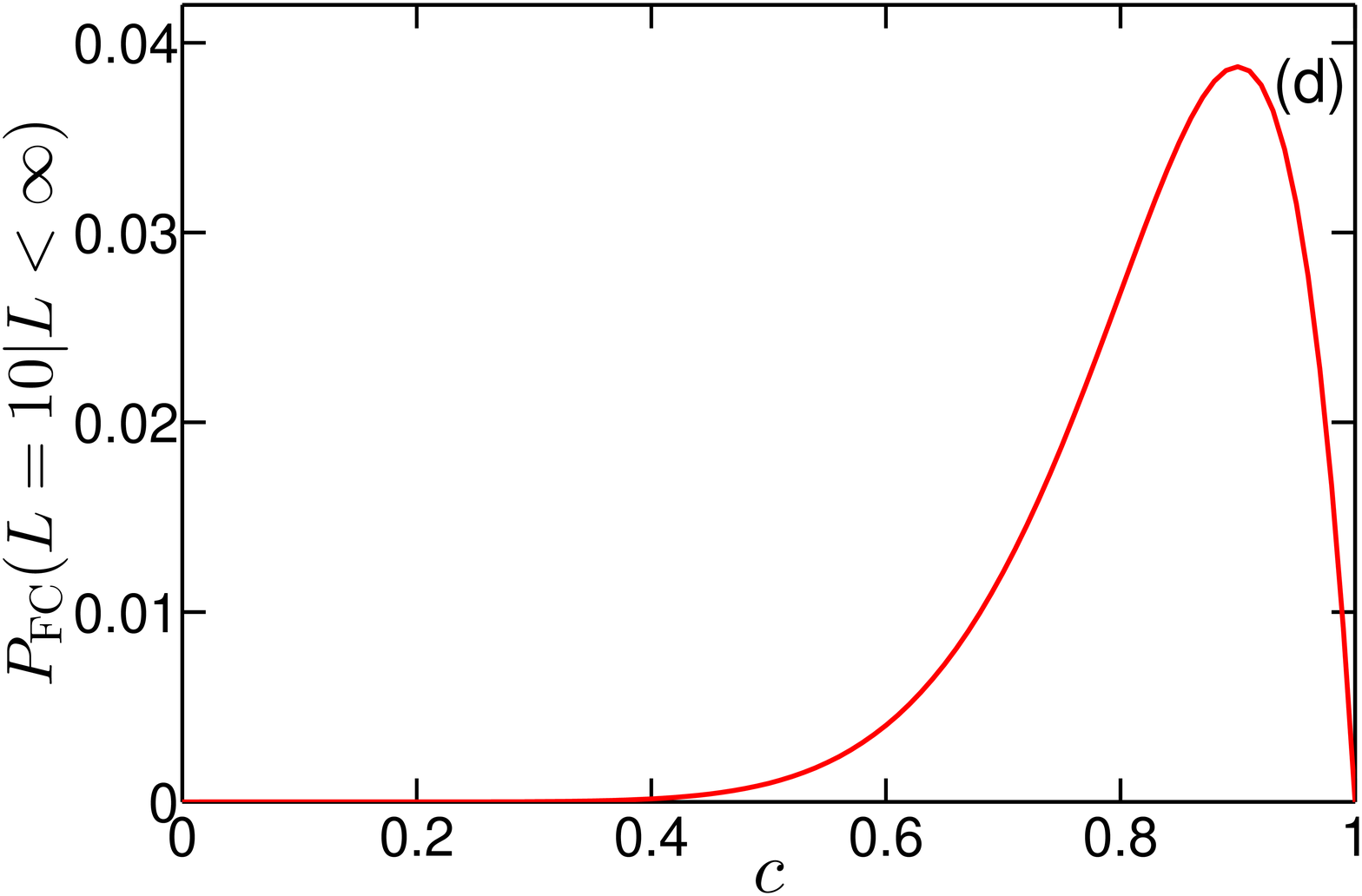}
\end{center}
\caption{
(Color online)
The probability $P_{\rm FC}(L=\ell | L < \infty)$ given by Eq. (\ref{eq:DSPL}),
is shown as a function of the mean degree, $c$, for $\ell=1, 2, 5$ and $10$.
The probability $P_{\rm FC}(L=1 | L<\infty)$ is a monotonically decreasing function of $c$.
For $\ell \ge 2$, the probability
$P_{\rm FC}(L=\ell | L < \infty)$ exhibits a peak at $c= 1 - 1/\ell$, which is the value of $c$
at which the probability of two random nodes to be at a distance $\ell$
from each other is maximal.
}
\label{fig:8}
\end{figure}

In Fig. \ref{fig:8} we present
the probability $P_{\rm FC}(L=\ell | L < \infty)$, given by Eq. (\ref{eq:DSPL}),
as a function of the mean degree, $c$, for $\ell=1, 2, 5$ and $10$.
The probability $P_{\rm FC}(L=1 | L<\infty)$ is a monotonically decreasing function of $c$.
This is due to the fact that for very low values of $c$ most of the components
consisting of two or more nodes are dimers and their fraction decreases
as $c$ is increased.
For $\ell \ge 2$, the probability
$P_{\rm FC}(L=\ell | L<\infty)$ vanishes at $c=0$ and $c=1$.
It increases for low values of $c$, reaches a peak and then starts to decrease.
For each value of $\ell \ge 2$, the peak of $P_{\rm FC}(L=\ell | L<\infty)$ is located
at $c= 1 - 1/\ell$, reflecting the appearance of longer paths 
as $c$ is increased.

It is also interesting to consider the conditional probabilities 
$P_{\rm FC}(L=\ell | L<\infty, K=k)$ 
and 
$P_{\rm FC}(L=\ell | L<\infty, K=k,K'=k')$,
between random
pairs of nodes that reside on the same finite component, 
under the condition that the degrees of one
or both nodes are specified, respectively.
In supercritical networks, the paths between nodes of high degrees tend to 
shorter than between nodes of low degrees. This is due to the fact that higher
degrees open more paths between the nodes, increasing the probability of
short paths to emerge. 
The situation in subcritical networks is completely different.
Any pair of nodes, $i$ and $j$, that reside on the same component are
connected by a single path. 
Such path goes through one neighbor of $i$ and one neighbor of $j$.
Therefore, the statistics of the path lengths
between pairs of nodes that reside on the same component
in subcritical ER networks do not depend
on the degrees of these nodes.
As a result, the conditional DSPLs satisfy

\begin{equation}
P_{\rm FC}(L=\ell | L<\infty, K=k) = (1-c) c^{\ell-1},
\end{equation}

\noindent
regardless of the value of $k$, and

\begin{equation}
P_{\rm FC}(L=\ell | L<\infty, K=k, K'=k') = (1-c) c^{\ell-1},
\end{equation}

\noindent
regardless of the values of $k$ and $k'$.
It is worth pointing out, however, that the probability that a random node
of a specified degree, $k$, and another random node of an unspecified degree
reside on the same component is dependent on the degree, $k$.
Using the results of Appendix B it can be shown that

\begin{equation}
P_{\rm FC}(L < \infty | K=k) = \frac{k}{(1-c)N}.
\end{equation}

\noindent
Similarly, it can be shown that the probability that a random node of
degree $k$ and another random node of degree $k'$ reside on the same
component is given by

\begin{equation}
P_{\rm FC}(L< \infty | K=k, K'=k') = \frac{k k'}{c(1-c)N}.
\end{equation}

\begin{figure}
\begin{center}
\includegraphics[width=12cm]{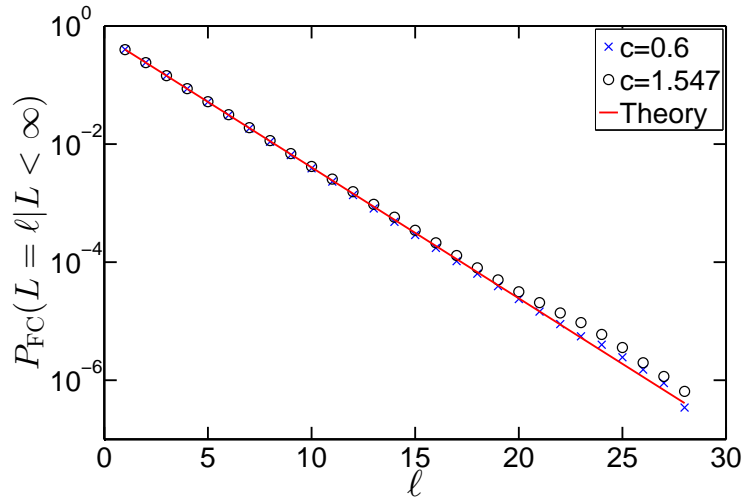}
\caption{
(Color online)
Numerical results
for the DSPL
on the finite components of an 
${\rm ER}[N,c/(N-1)]$ 
network with $N=10^4$ 
and $c=1.547$,
above percolation (circles)
and on its dual network,
${\rm ER}[N^{\prime},c^{\prime}/(N^{\prime}-1)]$
where $N^{\prime} = 3882$ [obtained from Eq. (\ref{eq:Np})]
and $c^{\prime}=0.6$ [obtained from Eq. (\ref{eq:cp})],
below percolation
($\times$),
which are essentially identical
and in excellent agreement with the theoretical results
(solid line),
obtained from Eq. (\ref{eq:DSPL}).
}
\label{fig:9}
\end{center}
\end{figure}

The DSPL of Eq. (\ref{eq:DSPL}) applies not only for subcritical ER networks
but also for the finite components of supercritical ER networks.
According to the duality relations, given a supercritical ER network of $N$
nodes and mean degree $c>1$, the subnetwork that consists of the
finite components is a subcritical ER network of size

\begin{equation}
N^{\prime} = N f(c),
\label{eq:Np}
\end{equation}

\noindent
and mean degree

\begin{equation}
c^{\prime} = c f(c),
\label{eq:cp}
\end{equation}

\noindent
where
$f(c) = - {\mathcal{W}( -c e^{-c} )}/{c}$
is the fraction of nodes in the supercritical network that reside
on the finite components and $c^{\prime}<1$.
In Fig. 
\ref{fig:9}
we present numerical results for the DSPL of the finite components of a supercritical
ER network of $N=10^4$ nodes and $c=1.547$ (circles).
The results are found to be in very good agreement with numerical
results for its dual network, which consists of $N' = 3882$ nodes and $c'=0.6$
($\times$) and with the analytical results for an asymptotic subcritical ER network
with $c=0.6$ (solid line), obtained from Eq. (\ref{eq:DSPL}).

\section{The mean and variance of the DSPL}

The moments of the DSPL provide useful information
about the large scale structure of the network. 
The first and second moments are
of particular importance. The first moment, 
$\langle L \rangle_{\rm FC}$
provides the mean distance.
The width of the DSPL is characterized by the
variance, 
${\rm Var}(L) = \langle L^2 \rangle - \langle L \rangle^2$,
where $\langle L^2 \rangle$ 
is the second moment.

The $n^{\rm th}$ moment of the DSPL over all 
trees of size $s$ and topology $\tau$ is given by

\begin{equation}
{\mathbb E}[L^n |\tau; L<\infty, S = s] = 
\sum_{\ell=1}^{s-1} \ell^n P_{\rm FC}(L=\ell |\tau; L<\infty,S=s),
\label{eq:E[L|tau_s]}
\end{equation}

\noindent
where
$P_{\rm FC}(L=\ell | \tau; L<\infty,S=s)$
is given by Eq. (\ref{eq:dspltaus}).
The 
$n^{\rm th}$ moment of the DSPL
over trees of all topologies, which consist of $s$
nodes is given by

\begin{equation}
{\mathbb E}[L^n | S = s] = 
\frac{
\sum\limits_{ \{ \tau | s \}} X(\tau) W(\tau;s) {\mathbb E}[L^n |\tau; L<\infty, S = s]
}{
\sum\limits_{ \{ \tau | s \}} X(\tau) W(\tau;s)
},
\label{eq:E[L|Seqs]}
\end{equation}

\noindent
where
${\mathbb E}[L |\tau; L<\infty, S = s]$
is given by Eq. (\ref{eq:E[L|tau_s]}).

The results for the first two moments,
${\mathbb E}[L|S=s]$
and
${\mathbb E}[L^2|S=s]$,
and for the variance
${\rm Var}[L|S=s] = {\mathbb E}[L^2|S=s] - ({\mathbb E}[L|S=s])^2$,
for $s=2,3,\dots,10$
are shown in Table
\ref{table:Pleqs}.

The $n^{\rm th}$ moment of the DSPL over all trees
that consist of up to $s$ nodes is given by

\begin{equation}
{\mathbb E}[L^n | S \le s] = 
\frac{ \sum\limits_{s'=2}^s \binom{s'}{2} P_{\rm FC}(S=s') {\mathbb E}[L^n | S=s']}
{ \sum\limits_{s'=2}^s \binom{s'}{2} P_{\rm FC}(S=s') },
\label{eq:E[L|Sles]}
\end{equation}

\noindent
where
$P_{\rm FC}(S=s')$
is given by Eq. (\ref{eq:Ps})
and
${\mathbb E}[L^n | S=s']$
is given by Eq. (\ref{eq:E[L|Seqs]}).
Performing the summation over all tree topologies up to size $s$
provides exact analytical expressions for the moments of the DSPL
over these trees.
The results for 
${\mathbb E}[L | S \le s']$
and
${\mathbb E}[L^2 | S \le s']$
For small trees of sizes
$s=2,3,\dots,6$
are shown in Table \ref{table:Plles}.
In the limit of large $s$, the mean distance ${\mathbb E}[L | S \le s]$
converges towards the asymptotic result, which is given by

\begin{equation}
\langle L \rangle_{\rm FC} = \frac{1}{1-c}.
\label{eq:ellmean}
\end{equation}

\begin{figure}
\begin{center}
\includegraphics[width=12cm]{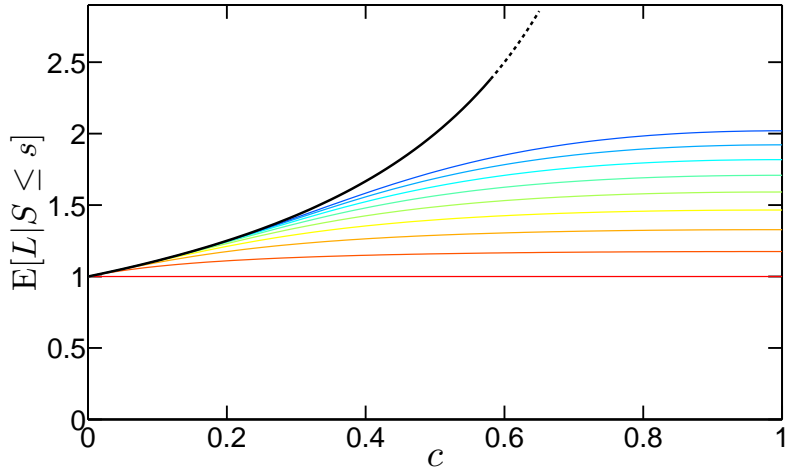}
\end{center}
\caption{
(Color online)
The mean distance
${\mathbb E}[L | S \le s]$,
over all tree topologies of sizes smaller or equal to $s$,
as a function of $c$,
for $s=2,3,\dots,10$
(solid lines), from bottom to top, respectively.
The thick solid line shows the asymptotic result,
$\langle L \rangle_{\rm FC}$.
}
\label{fig:10}
\end{figure}

\noindent
In Fig. \ref{fig:10} we present
the mean distances
${\mathbb E}[L | S \le s]$
(solid lines)
over all tree topologies of sizes smaller or equal to $s$,
as a function of $c$,
for $s=2,3,\dots,10$
(from bottom to top, respectively).
The thick solid line shows the asymptotic result,
$\langle L \rangle_{\rm FC}$,
given by Eq. (\ref{eq:ellmean}).
Clearly, as the tree size,
$s$, is increased, ${\mathbb E}[L | S \le s]$ approaches the asymptotic
result. As expected, for $c \ll 1$ the convergence is fast, but
as $c$ approaches the 
percolation threshold, the asymptotic result diverges 
and the convergence slows down. 

Using Eq. (\ref{eq:E[L|Sles]})
one can obtain exact analytical expressions for the second moment
of the DSPL over all trees of size $S \le s$. The results for small
trees that consist of up to $s=2,3,\dots,6$ nodes
are shown in Table \ref{table:Plles}.
In the limit of large $s$, the second moment ${\mathbb E}[L^2 | S \le s]$
converges towards the asymptotic result, which is given by

\begin{equation}
\langle L^2 \rangle_{\rm FC} = \frac{1+c}{(1-c)^2}.
\label{eq:ell2mean}
\end{equation}

The variance of the DSPL over all trees that consist of up
to $s$ nodes is given by

\begin{equation}
{\rm Var}[L | S \le s]= {\mathbb E}[L^2 | S \le s] - ({\mathbb E}[L | S \le s])^2.
\end{equation}

\noindent
Using the results presented 
in Table \ref{table:Plles}
for the first and second moments of 
the degree distributions over small trees
of sizes $s=2,3,\dots,5$,
we obtain

\begin{eqnarray}
{\rm Var}[L | S \le 2] &=&
0
\nonumber \\
{\rm Var}[L | S \le 3] &=&
\frac{\eta +  2 \eta^2 }{(1 + 3 \eta )^2}
\nonumber \\
{\rm Var}[L | S \le 4] &=&
\frac{ \eta +9 \eta^2+  19 \eta^3 + 31 \eta^4}
{\left( 1  + 3 \eta  +  8 \eta ^2  \right)^2}
\nonumber \\
{\rm Var}[L | S \le 5] &=&
\frac{ 36 \eta +324 \eta^2 +  1854 \eta^3  +4044\eta^4+ 7950 \eta^5   +   12054 \eta^6 }
{\left( 6  +18 \eta+  48 \eta ^2 +  125 \eta ^3   \right)^2}.
\end{eqnarray}

\noindent
In the limit of large $s$, the variance ${\rm Var}[L | S \le s]$
converges towards the asymptotic result, 
$\sigma^2_{L, {\rm FC}} ={\rm Var}(L)$,
where $\sigma_{L, {\rm FC}}$ is the standard
deviation of the DSPL over all the finite components.
The asymptotic variance is given by

\begin{equation}
{\rm Var}(L) =  \langle L^2 \rangle_{\rm FC} - \langle L \rangle_{\rm FC}^2.
\end{equation}

\noindent
Using Eqs. (\ref{eq:ellmean}) and (\ref{eq:ell2mean}) 
we find that

\begin{equation}
\sigma^2_{L,{\rm FC}} =
{\rm Var}(L) = 
\frac{c}{(1-c)^2}.
\end{equation}

\begin{figure}
\begin{center}
\includegraphics[width=11cm]{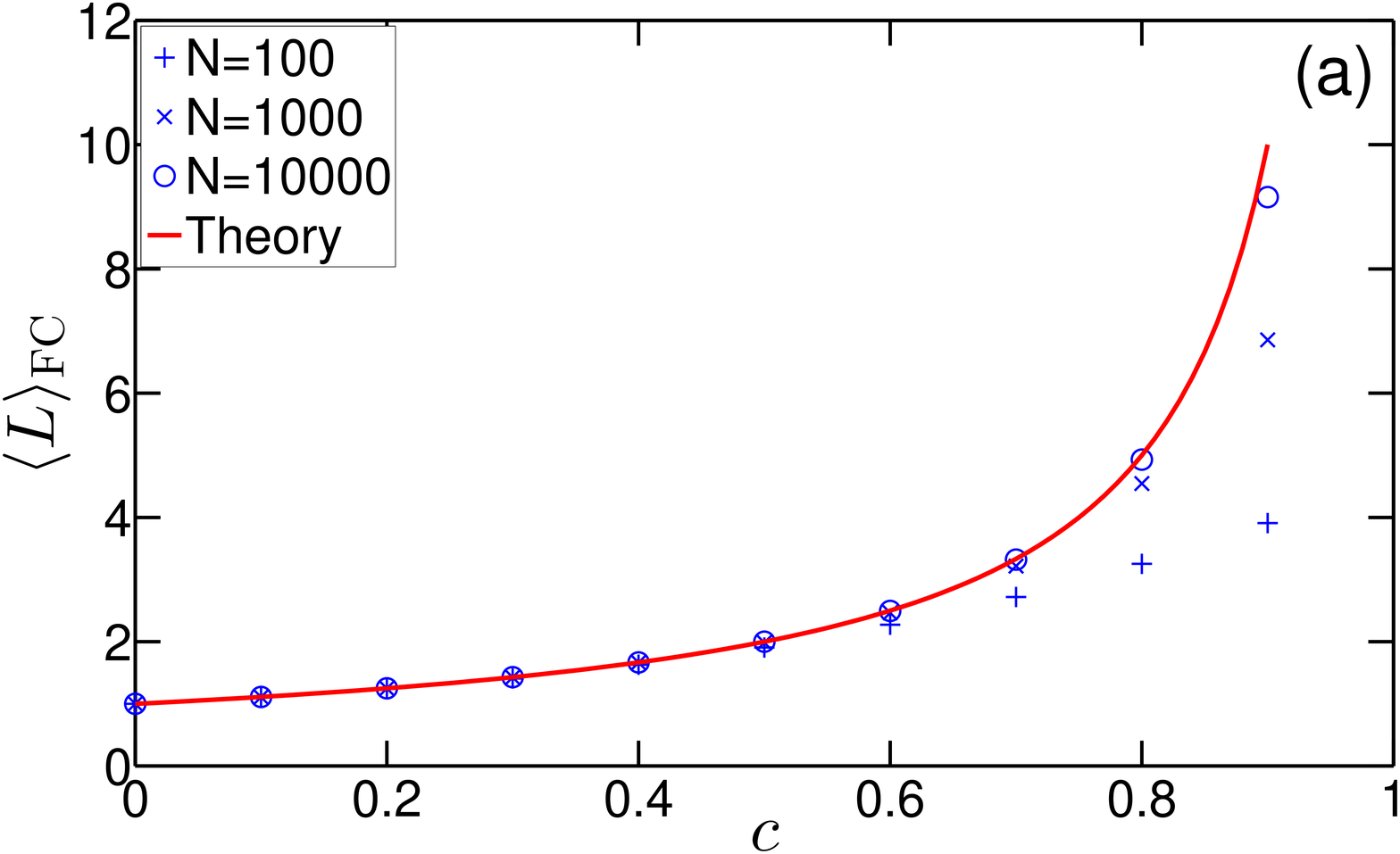}
\includegraphics[width=11cm]{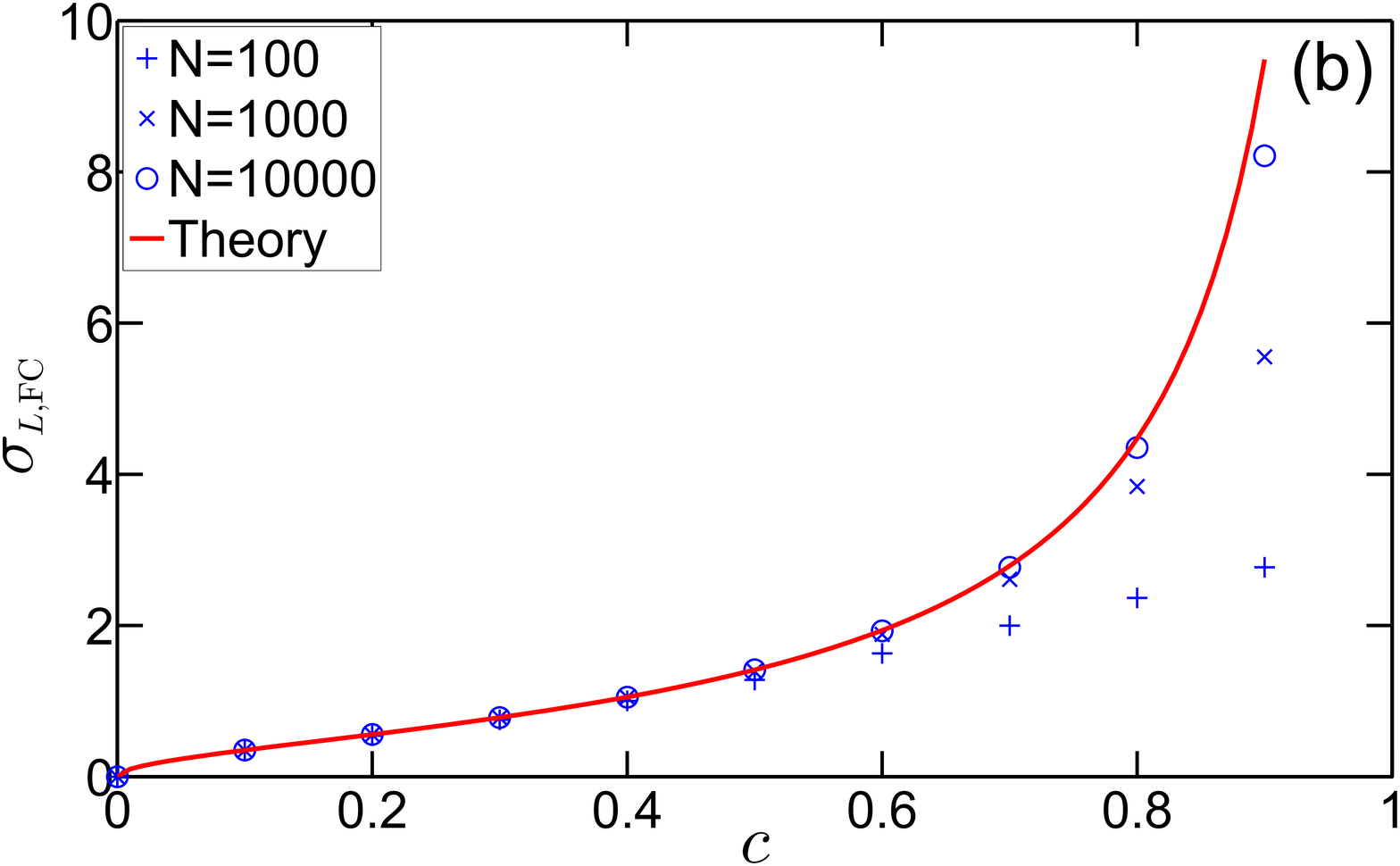}
\end{center}
\caption{
(Color online)
The mean, $\langle L \rangle_{\rm FC}$ (a), 
and the standard deviation, $\sigma_{L,{\rm FC}}$ (b), of
the DSPL of a subcritical
ER network vs. the mean degree, $c$.
The numerical results 
(symbols)
for 
$N=10^2$, $10^3$ and $10^4$ 
clearly converge towards the
analytical results 
(solid lines).
}
\label{fig:11}
\end{figure}

In Fig. 
\ref{fig:11}
we present the mean distance, $\langle L \rangle_{\rm FC}$ (a),
and the standard deviation $\sigma_{L,{\rm FC}}$ (b),
vs. the mean
degree, $c$. 
The theoretical results (solid lines) 
correspond to the asymptotic limit.
The numerical results,
obtained for 
$N=10^2$ ($+$), $10^3$ ($\times$) and $10^4$ ($\circ$)
are found to converge 
towards the theoretical results 
as the network size is increased.

\section{Discussion}

Apart from the shortest path length, random networks exhibit
other distance measures such as the resistance distance
\cite{Deza2009,Derrida1982,Bapat2014}.
The resistance distance, $r_{ij}$, 
between a pair of nodes, $i$ and $j$ 
is the electrical resistance between them under conditions in
which each edge in the network represents a resistor of $1$ Ohm.
Unlike the shortest path length, the resistance distance depends
on all the paths between $i$ and $j$, which often merge and
split along the way. 
It can be evaluated using the standard rules under which the
total resistance of resistors connected in series is the sum of their individual resistance values,
while the total resistance of resistors connected in parallel is the 
reciprocal of the sum of the reciprocals of the individual resistance values.
It was shown that 
the resistance distance between nodes $i$ and $j$ in a network can
be decomposed in terms of the eigenvalues and eigenvectors of the
normalized Laplacian matrix of the network
\cite{Chen2007,Yang2013}.
In order to utilize this result for the calculation of
the full distribution of resistance distances, 
$P(R=r)$,
in an ensemble of supercritical ER networks, 
one will need to obtain the full statistics of the spectral properties 
of the Laplacian matrix over the ensemble, which
is expected to be a difficult task.
For subcritical ER networks the situation is simpler.
Since the finite components in subcritical networks are trees, the shortest path between
a pair of nodes, $i$ and $j$, is in fact the only path between them.
As a result, the resistance distance between $i$ and $j$ is equal to the
shortest path length between them.
This means that the results presented in this paper provide not only the
distribution of shortest path lengths in subcritical ER networks but also
the distribution of resistance distances in these networks, 
which is given by

\begin{equation}
P_{\rm FC}(R=r | R < \infty) = (1-c)c^{r-1},
\end{equation}

\noindent
where $r$ takes integer values.

Another distance measure between nodes in random networks is
the mean first passage time, $t_{ij}$, of a random walk (RW) starting
from node $i$ and reaching node $j$
\cite{Sood2005,Debacco2015}. 
Unlike the shortest path length,
the mean first passage time is not
symmetric, namely $t_{ij} \ne t_{ji}$.
Since a RW may wander through side branches,
the mean first passage time
cannot be shorter than the shortest path, namely
$t_{ij} \ge \ell_{ij}$.
However, apart from this inequality,
there is no simple way to connect between these two quantities.
Therefore, numerical simulations will be suitable here.
Using specific large-deviation algorithms 
\cite{Hartmann2002,Hartmann2014}, 
it is possible, in
principle, to sample
the distributions over its full support, i.e. down to very small
probabilities like $10^{-100}$. Such approaches have been already applied
to obtain distributions of several properties of random graphs, e.g.,
the distribution of the number of components 
\cite{Engel2004}, 
the distribution of the
size of the largest component 
\cite{Hartmann2011}, 
the distribution of the 2-core
size 
\cite{Hartmann2017b}
or the distribution of the diameters 
\cite{Hartmann2017}.

Unlike RWs, which would eventually visit all the nodes in the component
on which they reside, the paths of self avoiding walks (SAWs) terminate
once they enter a leaf node
\cite{Tishby2016}.
Therefore, an SAW starting from node $i$ does not necessarily reach
node $j$ even if they reside on the same component. 
However, in case it reaches node $j$ its first passage time is
equal to the shortest path length between $i$ and $j$.
Therefore, the distribution of first passage times,
$P_{\rm SAW}(T=t | T<\infty)$,
of SAWs between pairs
of nodes that reside on the same component
satisfies
$P_{\rm SAW}(T=t | T < \infty) = P_{\rm FC}(L=\ell | L<\infty)$.

The DSPL of subcritical ER networks is also relevant to the
study of epidemic spreading on supercritical ER networks.
Consider a supercritical ER network with mean degree $c>1$. 
An epidemic starts from a random node, $i$, and propagates through
the shell structure around this node. 
The time is discrete, so each node that is infected
at time $t$ may infect each one of its neighbors at time $t+1$, 
with probability $p'$. The node that was infected at time $t$ 
recovers at time $t+1$ and becomes immune.

The expectation value of the number of nodes infected
by node $i$ in the first time step is given by 
$c' = c p'$.
In case that $c'<1$, the statistical properties of the 
components formed by such epidemic are similar to the statistical
properties of the tree components in a subcritical network with
mean degree $c'$. More precisely, the size distribution of components formed by the
epidemic follows the distribution 
$P_{\rm FC}(S=s)$
of component sizes on which a random
node resides, given by 
Eq. (\ref{eq:Pstilde}).
This property represents some kind of invariance, the distribution of 
epidemic sizes depends only on
the value of the product $c'=c p'$ 
rather than the values of $c$ and $p'$ alone. 
The DSPL, 
$P_{\rm FC}(L=\ell | L<\infty)$ 
represents the temporal propagation
of a typical epidemic, namely the probability that a node that was
infected by an epidemic got infected $ell$ time steps after the epidemic
started.

Using extreme value statistics it may be possible to obtain analytical
results for the distributions of radii and diameters over all the
tree topologies.
For networks that satisfy duality relations, it will be possible to 
obtain the DSPL on the finite components in the supercritical regime.
Combining the results with the DSPL on the giant component will yield 
the overall DSPL of the supercritical network.
The detailed understanding of the DSPL in terms of the topological
expansion is expected to be useful in the study of dynamical processes
such as epidemic spreading.
Since epidemic spreading and many other real-world dynamical processes 
take place on networks that are different from ER networks, it
will be interesting to apply the topological expansion presented here
to the analysis of the DSPL in a broader class of subcritical 
random networks.
In particular, extending this approach to configuration model networks will
provide the DSPL of subcritical random networks with any desired degree
distribution. To this end, one will need to derive an equation for the size distribution
of the finite tree components, $P_{\rm FC}(S=s)$ in a configuration model network
with a given degree distribution, $P(K=k)$. 
The weights, $W(\tau;s)$, of the different tree topologies, $\tau$, which
consist of $s$ nodes, in a configuration model network
are expected to depend on the degree distribution.
Therefore, one will need to derive an equation for $W(\tau,s)$ in terms
of $P(K=k)$. Once the weights become available, the counting of the shortest
paths follows the same procedure used in the ER case.
It will also be interesting to apply the topological expansion to 
edge-independent, inhomogeneous random graphs
\cite{Chung2003,Bollobas2007,Lu2013}.
This family of network models provides a generalization of the ER network,
in which the probability $p$ 
is replaced by a random $N \times N$ matrix, ${\mathcal P}$, in which the matrix element
${\mathcal P}_{ij}$ is the probability that nodes $i$ and $j$ are connected by an edge.
As a result, each node, $i$, exhibits unique statistical properties
that depend on ${\mathcal P}_{ij}$, $j=1,2,\dots,N$,
leading to non-Poissonian degree distributions,
as in the case of configuration model networks.
To apply the topological expansion to inhomogeneous random graphs
one will need to perform an additional summation over the distribution
of the matrix elements of ${\mathcal P}$.

\section{Summary}

We have developed a topological expansion methodology
for the analysis of subcritical random networks.
The expansion is based on the fact that such networks
are fragmented into finite tree components, which can be
classified systematically by their sizes and topologies.
Using this approach we performed a systematic calculation of
the degree distribution,
$P_{\rm FC}(K=k | S \le s)$,
and the DSPL 
$P_{\rm FC}(L=\ell | L<\infty,S \le s)$,
over all components whose size is smaller or equal to $s$,
in subcritical ER networks.
Taking the large $s$ limit, we obtained an exact 
asymptotic formula for the DSPL over all pairs of nodes
that reside on the same component,
which takes the form

\begin{equation}
P_{\rm FC}(L=\ell | L<\infty)=(1-c)c^{\ell-1}.
\label{eq:PLLs}
\end{equation}

\noindent
This remarkably simple asymptotic result is obtained only when the contributions
of the tree components of all sizes and topologies are taken into account.
Such mean-field like results are normally expected to represent the shell structure around a
typical node. However, in subcritical networks there is no typical node because the
shell structure strongly depends on the size and topology of the tree component
in which each node resides as well as on its location in that component.

From the degree distribution and the DSPL, we obtained analytical results for the
mean degree, the variance of the degree distribution,
the mean distance and the variance of the DSPL over all
components whose size is smaller or equal to $s$.
Taking the large $s$ limit, we found that
the mean path length between all pairs of nodes that reside
on the same component is given by

\begin{equation}
\langle L \rangle_{\rm FC} = \frac{1}{1-c}. 
\label{eq:AvLs}
\end{equation}

\noindent
As the percolation threshold is approached from below, 
at $c \rightarrow 1^{-}$,
the mean distance diverges as
$\langle L \rangle_{\rm FC} \sim (1-c)^{-\alpha}$,
where the exponent  
$\alpha=1$.
From the duality relations between a subcritical ER network
and the finite components in a corresponding supercritical ER network,
it is found that the same exponent,
$\alpha=1$, appears also above the transition.

\appendix

\section{The distribution of tree sizes}

In this Appendix we review some useful results on the distribution
of tree sizes in subcritical ER networks.
Consider a subcritical ER network of $N$ nodes with mean degree $c<1$.
The expectation value of the number of trees of $s$ nodes in such
network is denoted by $T_s^N$.
Using the theory of branching processes, it was shown that 
$T_s^N$ is given by
\cite{Bollobas2001,Durrett2007}

\begin{equation}
T_s^N = N { \binom{N}{s} } s^{s-2} \left( \frac{c}{N} \right)^{s-1} 
\left(1-\frac{c}{N} \right)^{ { \binom{s}{2} } -(s-1)}
\left(1-\frac{c}{N} \right)^{s(N-s)} ,
\end{equation}

\noindent
where the binomial coefficient accounts for the number of ways
to pick $s$ nodes out of $N$ in order to form a component of size $s$
and the factor of $s^{s-2}$ is the number of distinct tree structures
that can be constructed from $s$ distinguishable nodes
\cite{Cayley1889}.
The factor of $(c/N)^{s-1}$ accounts for the probability that the
$s$ nodes of the component will be connected by $s-1$ edges. 
The next term is the probability that there are no other edges
connecting pairs of nodes in the component,
while the last term is the probability that 
there are no edges connecting nodes in the components with nodes
in the rest of the network.
For $s \ll N$ one can approximate the binomial coefficient by $N^s/s!$
and obtain

\begin{equation}
T_s^N = N  \frac{ s^{s-2} c^{s-1} }{s!}
\left(1-\frac{c}{N} \right)^{s(N-s) + { \binom{s}{2} } -(s-1)}.
\label{eq:Tsf}
\end{equation}

\noindent
Since we consider subcritical ER networks, for which $c < 1$,
unless the network is extremely small the condition
$c \ll N$ is satisfied. 
Therefore, one can approximate
the last term 
in Eq. (\ref{eq:Tsf})
by an exponential, and obtain

\begin{equation}
T_s^N = N  \frac{ s^{s-2} c^{s-1} e^{-cs} }{s!} 
\exp \left[\frac{c (s^2+3s-2)}{2N} \right].
\label{eq:Tsf2}
\end{equation}

\noindent
Finally, in the asymptotic limit of 
$N \rightarrow \infty$,
the exponential converges towards $1$ and
the expression for the expected number of tree components 
of $s$ nodes is reduced to
\cite{Bollobas2001,Durrett2007}

\begin{equation}
T_s^N  \simeq N  \frac{ s^{s-2}{c^{s - 1}}e^{-c s} }{ s! }.
\label{eq:T_sinf}
\end{equation}

\noindent
In the limit of large $s$, one can use the Stirling approximation 
and obtain

\begin{equation}
T_s^N \simeq \frac{N}{\sqrt{2 \pi} c} \frac{ e^{-s/s_{\rm max}} }{ s^{5/2} },
\label{eq:T_sinf2}
\end{equation}

\noindent
where the cutoff parameter $s_{\rm max}$ is given by

\begin{equation}
s_{\rm max} = \frac{1}{\ln \left( \frac{1}{c e^{1-c}} \right)}.
\end{equation}

As the percolation threshold is approached from below, for
$c \rightarrow 1^{-}$, 
the cutoff parameter diverges, according to
$s_{\rm max} \sim 1/(1-c)^2$.
The expected number of trees of size $s$ per node,
obtained from Eq. (\ref{eq:T_sinf2})
scales like
$T_s^N/N  \propto s^{-\tau}$,
where $\tau=5/2$.
This is in agreement with the critical component size distribution on
regular lattices above the upper critical dimension
of $D=6$, where $\tau$ is the Fisher exponent
\cite{Stauffer2003},
exemplifying the connection between percolation transitions
on random networks and regular lattices of high dimensions.

The total number of tree components in a 
subcritical ER
network of $N$ nodes and $c<1$
is denoted by

\begin{equation}
N_T(c) = \sum_{s=1}^{N} T_s^N.
\end{equation}

\noindent
Carrying out the summation, 
using the expression for $T_s^N$
from Eq. (\ref{eq:T_sinf}),
we obtain that for $0 \le c \le 1$

\begin{equation}
N_T(c) =  \left( 1 - \frac{c}{2} \right) N,
\label{eq:numtrees}
\end{equation}

\noindent
namely $N_T(c)$ is a linear, monotonically decreasing function of
$c$, where $N_T(c=0)=N$ and $N_T(c=1)=N/2$.
The mean tree size is thus given by

\begin{equation}
\langle S \rangle_{\rm FC} = \frac{2}{2-c},
\end{equation}

\noindent
which does not diverge as $c$ approaches the percolation threshold.
Using  
Eqs. (\ref{eq:T_sinf}) and (\ref{eq:numtrees})
we can write down the distribution
of tree sizes, which takes the form

\begin{equation}
P_{\rm FC}(S=s) = \frac{ 2 s^{s-2} c^{s-1} e^{-c s} }{(2-c) s! }.
\label{eq:Ps}
\end{equation}

\noindent
In various processes on networks components are selected by
by drawing random nodes and choosing the components on
which they reside.
The probability that a randomly selected node resides on a
tree of size $s$ is given by 

\begin{equation}
\widetilde P_{\rm FC}(S=s) 
= \frac{s}{\langle S \rangle_{\rm FC}} P_{\rm FC}(S=s).
\label{eq:Pstilde}
\end{equation}

\noindent
The mean of this distribution is

\begin{equation}
\langle \widetilde S \rangle_{\rm FC} 
= \frac{\langle S^2 \rangle_{\rm FC}}{\langle S \rangle_{\rm FC}} 
= \frac{1}{1-c}.
\end{equation}

\noindent
Thus, as $c \rightarrow 1^{-}$, the mean tree size
on which a random node resides diverges.

Consider a random pair of nodes that reside on the same component.
The probability that they reside on a component of size $s$ is 
given by

\begin{equation}
\widehat P_{\rm FC}(S=s) 
= \frac{ \binom{s}{2} P_{\rm FC}(S=s) }{ \left\langle \binom{S}{2} \right\rangle_{\rm FC}}.
\end{equation}

\noindent
Evaluating the denominator we obtain

\begin{equation}
\left\langle \binom{S}{2} \right\rangle_{\rm FC} = \frac{c}{(1-c)(2-c)}.
\end{equation}

\noindent
The mean of $\widehat P_{\rm FC}(S=s)$ is found to be

\begin{equation}
\langle \widehat S \rangle_{\rm FC} = \frac{2-c}{(1-c)^2},
\end{equation}

\noindent
which diverges quadratically as $c \rightarrow 1^{-}$.

\section{The probability that two random nodes reside on the same component}

In this Appendix we calculate the probability,
$P_{\rm FC}(L < \infty)$,
that two random nodes in a subcritical ER network
reside on the same component.
This probability is given by

\begin{equation}
P_{\rm FC}(L < \infty) = \frac{ \mathcal{L}(N,c) }{ { \binom{N}{2} } },
\label{eq:PLf}
\end{equation}

\noindent
where 
$\mathcal{L}(N,c)$
is the number of pairs of nodes that reside on the same
component. It is given by

\begin{equation}
\mathcal{L}(N,c) = \sum_{s \ge 1} { \binom{s}{2} } T_s^N,
\end{equation}

\noindent
where $T_s^N$ is the number of tree components of size $s$,
given by Eq. (\ref{eq:T_sinf}).
In order to evaluate this sum we use properties of the
Lambert $W$ function, denoted by
$\mathcal{W}(x)$
\cite{Olver2010}.
In particular, we use the implicit definition
(Eq. 4.13.1 in Ref.
\cite{Olver2010}):

\begin{equation}
\mathcal{W}(x) = x e^{- \mathcal{W}(x)}.
\end{equation}

\noindent
We also use the series expansion
(Eq. 4.13.5 in Ref.
\cite{Olver2010}):

\begin{equation}
\mathcal{W}(x) = - \sum_{s=1}^{\infty} \frac{s^{s-2}}{s!} s (-x)^s.
\label{eq:Wexp}
\end{equation}

\noindent
Using the series expansion of Eq.
(\ref{eq:Wexp})
it can be shown that

\begin{equation}
\sum_{s=1}^{\infty} { \binom{s}{2} } \frac{s^{s-2}}{s!}  (-x)^s =
\frac{1}{2}
\left[  \mathcal{W}(x) - x \frac{d}{dx}  \mathcal{W}(x)  \right] 
\label{eq:Wexp2}
\end{equation}

\noindent
Plugging in
Eq. 4.13.4 of Ref.
\cite{Olver2010},
which can be expressed in the form

\begin{equation}
\frac{d}{dx} \mathcal{W}(x) 
= \frac{\mathcal{W}(x)}{x \left[1+\mathcal{W}(x) \right]},
\end{equation}

\noindent
we obtain

\begin{equation}
\sum_{s=1}^{\infty} { \binom{s}{2} } \frac{s^{s-2}}{s!}  (-x)^s =
\frac{ [\mathcal{W}(x)]^2 }{2\left[1+\mathcal{W}(x) \right]}.
\label{eq:Wexp3}
\end{equation}

\noindent 
Plugging in $x=-c e^{-c}$,
multiplying by $N/c$
and using the representation of $T_s^N$ in Eq. (\ref{eq:T_sinf}),
we obtain for the left-hand side

\begin{equation}
\frac{N}{c} \sum_{s=1}^{\infty} { \binom{s}{2} } \frac{s^{s-2}}{s!}  (ce^{-c})^s 
= \sum_{s=1}^{\infty} { \binom{s}{2} } N\frac{s^{s-2} c^{s-1}e^{-cs}}{s!}  
= \sum_{s=1}^{\infty} { \binom{s}{2} } T_s^N\,,
\end{equation}

\noindent
which is the quantity we want, 
for finite values of $N$. Therefore,
we obtain

\begin{equation}
\mathcal{L}(N,c) = 
\left( \frac{N}{2c} \right)
\frac{ [\mathcal{W}(-c e^{-c})]^2}{ 1 + \mathcal{W}(-c e^{-c})}.
\end{equation}

\noindent
For $0<c<1$ it can be shown that
$\mathcal{W}(-c e^{-c}) = -c$,
and thus

\begin{equation}
\mathcal{L}(N,c) = \frac{N c}{2(1-c)}.
\end{equation}

\noindent
Using Eq. (\ref{eq:PLf}) we find that in the asymptotic limit, 
$N \rightarrow \infty$,
the probability that two randomly selected nodes in the network 
reside on the same component is given by

\begin{equation}
P_{\rm FC}(L < \infty) =
\frac{c}{(1-c)(N-1)} \simeq 
\frac{c}{(1-c)N}. 
\end{equation}

\clearpage
\newpage

\begin{table}
\caption{The different degree distributions and DSPLs 
for the finite components (FC) of subcritical ER networks
and the equations which are used to evaluate them.}
\begin{tabular}{| l  | l | l | }
\hline \hline
{\bf Distribution}  & {\bf Equation} & {\bf Description}    \\ 
\hline \hline 
$P_{\rm FC}(K=k |\tau; S=s)$   
&  
{\bf Eq. (\ref{eq:Pktaus}) }   
& 
Degree distribution over all trees of $s$ nodes and topology $\tau$
\\ \hline
$P_{\rm FC}(K=k | S=s)$   
&  
{\bf Eq. (\ref{eq:P_k2}) }   
& 
Degree distribution over all trees of $s$ nodes
\\ \hline
$P_{\rm FC}(K=k | S \le s)$   
&  
{\bf Eq. (\ref{eq:dspk3}) } 
&  
Degree distribution over all trees of up to $s$ nodes 
\\ \hline
$P_{\rm FC}(K=k)$   
&  
{\bf Eq. (\ref{eq:poisson_k>0}) } 
&  
Degree distribution over all trees 
\\ \hline
\hline
${\mathbb E}[K | \tau; S=s]$
&
{\bf Eq. (\ref{eq:E[K|tau_s]}) }
&
Mean degree over all trees of $s$ nodes and topology $\tau$
\\ \hline
${\mathbb E}[K | S=s]$
&
{\bf Eq. (\ref{eq:E[K|s]}) }
&
Mean degree over all trees of $s$ nodes 
\\ \hline
${\mathbb E}[K | S \le s]$
&
{\bf Eq. (\ref{eq:E[K|Sles]}) }
&
Mean degree over all trees of up to $s$ nodes 
\\ \hline
$\langle K \rangle_{\rm FC}$
&
{\bf Eq. (\ref{eq:<K>}) }
&
Mean degree over all trees 
\\ \hline
\hline
$P_{\rm FC}(L = \ell |\tau; L<\infty,S=s)$   
&  
{\bf Eq. (\ref{eq:dspltaus}) }   
& 
DSPL over all trees of $s$ nodes and topology $\tau$ 
\\ \hline
$P_{\rm FC}(L = \ell | L<\infty,S=s)$   
&  
{\bf Eq. (\ref{eq:dspl2}) }   
& 
DSPL over all trees of $s$ nodes 
\\ \hline
$P_{\rm FC}(L = \ell | L<\infty, S \le s )$   
&  
{\bf Eq. (\ref{eq:dspl3})} 
&  
DSPL over all trees of up to $s$ nodes
\\ \hline
$P_{\rm FC}(L = \ell | L<\infty)$   
&  
{\bf Eq. (\ref{eq:DSPL})} 
&  
DSPL over all trees 
\\ \hline
\hline
${\mathbb E}[L | \tau; S =s]$
&
{\bf Eq. (\ref{eq:E[L|tau_s]})} 
&
Mean distance over all trees of $s$ nodes and topology $\tau$
\\ \hline
${\mathbb E}[L |  S =s]$
&
{\bf Eq. (\ref{eq:E[L|Seqs]})} 
&
Mean distance over all trees of $s$ nodes 
\\ \hline
${\mathbb E}[L |  S  \le s]$
&
{\bf Eq. (\ref{eq:E[L|Sles]})} 
&
Mean distance over all trees of up to $s$ nodes 
\\ \hline
$\langle L \rangle_{\rm FC}$
&
{\bf Eq. (\ref{eq:ellmean})} 
&
Mean distance over all trees
\\ \hline
\hline
\end{tabular}
\label{table:P}
\end{table}

\clearpage
\newpage

\begin{table}
\caption{
The conditional probabilities $P_{\rm FC}(K=k | S=s)$, that
a node selected randomly from all the tree components of size $s$,
in a subcritical ER network, will have a degree $k$.
Results are shown for $s=2,3,\dots,10$ nodes.
}
\vspace{0.4in}
\begin{tabular}{| l  | r  | r | r | r |r | r | r | r | r |}
\hline \hline
& $s = 2$
     &  $s = 3$  &  $s = 4$ & $s = 5$ & $s = 6$ & $s = 7$ & $s = 8$ & $s = 9$ & $s = 10$ \\ 
\hline \hline 
$P_{\rm FC}(K=1 | S=s)$ =
&
$1$
&
$\frac{2}{3}$
&
$\frac{ 9}{16}$
&
$\frac{ 64}{125}$
&
$\frac{625}{1296}$
&
$\frac{7776}{16807}$
&
$\frac{117649}{262144}$
&
$\frac{2097152}{4782969}$
&
$\frac{43046721}{100000000}$
\\
\hline
$P_{\rm FC}(K=2 | S=s) =$
& 
&
$\frac{1}{3}$
&
$\frac{3}{8}$
&
$\frac{48}{125}$
&
$\frac{125}{324}$
&
$\frac{6480}{16807}$
&
$\frac{50421}{131072}$
&
$\frac{1835008}{4782969}$
&
$\frac{4782969}{12500000}$
\\
\hline
$P_{\rm FC}(K=3 | S=s) =$
&
&
&
$\frac{1}{16}$
&
$\frac{12}{125}$
&
$\frac{25}{216}$
&
$\frac{2160}{16807}$
&
$\frac{36015}{262144}$
&
$\frac{229376}{1594323}$
&
$\frac{3720087}{25000000}$
\\
\hline
$P_{\rm FC}(K=4 | S=s) = $
&
&
&
&
$\frac{1}{125}$
&
$\frac{5}{324}$
&
$\frac{360}{16807}$
&
$\frac{1715}{65536}$
&
$\frac{143360}{4782969}$
&
$\frac{413343}{12500000}$
\\
\hline
$P_{\rm FC}(K=5 | S=s)=$
&
&
&
&
&
$\frac{1}{1296}$
&
$\frac{30}{16807}$
&
$\frac{735}{262144}$
&
$\frac{17920}{4782969}$
&
$\frac{45927}{10000000}$
\\
\hline
$P_{\rm FC}(K=6 | S=s)=$
&
&
&
&
&
&
$\frac{1}{16807}$
&
$\frac{21}{131072}$
&
$\frac{448}{1594323}$
&
$\frac{5103}{12500000}$
\\
\hline
$P_{\rm FC}(K=7 | S=s) =$
&
&
&
&
&
&
&
$\frac{1}{262144}$
&
$\frac{56}{4782969}$
&
$\frac{567}{25000000}$
\\
\hline
$P_{\rm FC}(K=8 | S=s)=$
&
&
&
&
&
&
&
&
$\frac{1}{4782969}$
&
$\frac{9}{12500000}$
\\
\hline
$P_{\rm FC}(K=9 | S=s)=$
&
&
&
&
&
&
&
&
&
$\frac{1}{100000000}$
\\
\hline
\hline
${\mathbb E}[K | S=s] =$
&
1
&
$\frac{4}{3}$
&
$\frac{3}{2}$
&
$\frac{8}{5}$
&
$\frac{5}{3}$
&
$\frac{12}{7}$
&
$\frac{7}{4}$
&
$\frac{16}{9}$
&
$\frac{9}{5}$
\\ \hline
${\mathbb E}[K^2 | S=s] =$
&
1
&
2
&
$\frac{21}{8}$
&
$\frac{76}{25}$
&
$\frac{10}{3}$
&
$\frac{174}{49}$
&
$\frac{119}{32}$
&
$\frac{104}{27}$
&
$\frac{99}{25}$
\\ \hline
${\rm Var}[K | S=s] =$
&
$0$
&
$\frac{2}{9}$
&
$\frac{3}{8}$
&
$\frac{12}{25}$
&
$\frac{5}{9}$
&
$\frac{30}{49}$
&
$\frac{21}{32}$
&
$\frac{56}{81}$
&
$\frac{18}{25}$
\\
\hline
\hline
\end{tabular}
\label{table:Pkeqs}
\vspace{0.4in}
\end{table}

\begin{table}
\caption{
The probabilities $P_{\rm FC}(K=k | 2 \le S \le s)$, that
a node selected randomly from all the tree components of sizes
in the range $2 \le S \le s$,
in a subcritical ER network, will have a degree $k$.
Results are shown for $s=2,3,\dots,10$ nodes.
}
\vspace{0.4in}
\begin{tabular}{| l  | r  | r | r | r |r | }
\hline \hline
& $s = 2$
     &  $s = 3$  &  $s = 4$ & $s = 5$ & $s = 6$  \\ 
\hline \hline 
$P_{\rm FC}(K=1 | 2 \le S \le s)$ =
&
$1$
&
$\frac{2+ 2 \eta}{2 + 3 \eta}$
&
$\frac{6 + 6 \eta + 9 \eta^2}{6 + 9 \eta + 16 \eta^2}$
&
$\frac{24 + 24 \eta + 36 \eta^2 + 64 \eta^3}{24 + 36 \eta + 64 \eta^2 + 125 \eta^3}$
&
$\frac{120 + 120 \eta + 180 \eta^2 + 320 \eta^3 + 625 \eta^4}
{120 + 180 \eta + 320 \eta^2 + 625 \eta^3 + 1296 \eta^4}$
\\
\hline
$P_{\rm FC}(K=2 | 2 \le S \le s) =$
& 
&
$\frac{\eta}{2 + 3 \eta}$
&
$\frac{3 \eta + 6 \eta^2}{6 + 9 \eta + 16 \eta^2}$
&
$\frac{12 \eta + 24 \eta^2 + 48 \eta^3}{24 + 36 \eta + 64 \eta^2 + 125 \eta^3}$
&
$\frac{60 \eta + 120 \eta^2 + 240 \eta^3 + 500 \eta^4}
{120 + 180 \eta + 320 \eta^2 + 625 \eta^3 + 1296 \eta^4}$
\\
\hline
$P_{\rm FC}(K=3 | 2 \le S \le s) =$
&
&
&
$\frac{\eta^2}{6 + 9 \eta + 16 \eta^2}$
&
$\frac{4 \eta^2 + 12 \eta^3}{24 + 36 \eta + 64 \eta^2 + 125 \eta^3}$
&
$\frac{20 \eta^2 + 60 \eta^3 + 150 \eta^4}
{120 + 180 \eta + 320 \eta^2 + 625 \eta^3 + 1296 \eta^4}$
\\
\hline
$P_{\rm FC}(K=4 | 2 \le S \le s) = $
&
&
&
&
$\frac{\eta^3}{24 + 36 \eta + 64 \eta^2 + 125 \eta^3}$
&
$\frac{5 \eta^3 + 20 \eta^4}{120 + 180 \eta + 320 \eta^2 + 625 \eta^3 + 1296 \eta^4}$
\\
\hline
$P_{\rm FC}(K=5 | 2 \le S \le s)=$
&
&
&
&
&
$\frac{\eta^4}{120 + 180 \eta + 320 \eta^2 + 625 \eta^3 + 1296 \eta^4}$
\\
\hline
\hline
${\mathbb E}[K | 2 \le S \le 2] =$
&
$1$
&
$\frac{2 + 4 \eta}{2 + 3 \eta}$
&
$\frac{6 + 12 \eta + 24 \eta^2}{6 + 9 \eta + 16 \eta^2}$
&
$\frac{24 + 48 \eta + 96 \eta^2 + 200 \eta^3}{24 + 36 \eta + 64 \eta^2 + 125 \eta^3}$
&
$\frac{120 + 240 \eta + 480 \eta^2 + 1000 \eta^3 + 2160 \eta^4}
{120 + 180 \eta + 320 \eta^2 + 625 \eta^3 + 1296 \eta^4}$
\\
\hline
${\mathbb E}[K^2 | 2 \le S \le s] =$
&
$1$
&
$\frac{2 + 6 \eta}{2 + 3 \eta}$
&
$\frac{6  +18 \eta +  42 \eta ^2}
{6 +9 \eta +  16 \eta ^2}$
&
$\frac{24 +72 \eta + 168 \eta ^2 + 380\eta^3}
{24 +36 \eta + 64 \eta ^2 + 125 \eta^3}$
&
$\frac{120  +360 \eta + 840 \eta^2 +1900\eta^3 + 4320 \eta^4}
{120 +180 \eta +320 \eta^2 +625 \eta^3 + 1296 \eta^4}$
\\
\hline
\hline
\end{tabular}
\label{table:Pkles}
\end{table}

\clearpage
\newpage

\begin{table}
\caption{The leading finite size correction terms,
${P_{\rm FC}(K=k | 2 \le S \le s)}/{\pi_{\rm FC}(K=1)} -1 \simeq q_{s,k} c^{s-k}$ 
of Eq. (\ref{eq:ddf}) for the
degree distribution over all the tree topologies with up to $s$ nodes
(except for the case of an isolated node).
The distribution
$\pi_{\rm FC}(K=k)$, given by Eq. (\ref{eq:poisson_k>0}),
is the degree distribution over the entire subcritical network, except
for the isolated nodes. As $s$ is increased the correction decreases
as $c^{s-1}$ and $P_{\rm FC}(K=k | 2 \le S \le s)$ 
converges towards $\pi_{\rm FC}(K=k)$.}
\vspace{0.4in}
\begin{tabular}{| l  | r  | r | r | r |r | r | r | r | r |}
\hline \hline
& $s = 2$
     &  $s = 3$  &  $s = 4$ & $s = 5$ & $s = 6$ & $s = 7$ & $s = 8$ & $s = 9$ & $s = 10$ \\ 
\hline \hline 
$ \frac{P_{\rm FC}(K=1 | 2 \le S \le s)}{\pi_{\rm FC}(K=1)} -1 \simeq  $   
& $ \frac{1}{2}c$
& $ \frac{7}{6}c^2$
& $ \frac{61}{24}c^3$
& $ \frac{671}{120}c^4 $
& $ \frac{9031}{720}c^5$
& $ \frac{3211}{112}c^6$
& $ \frac{2685817}{40320}c^7$
& $ \frac{56953279}{362880}c^8$
& $ \frac{1357947691}{3628800}c^9$
\\
\hline
$ \frac{P_{\rm FC}(K=2|2 \le S \le s)}{\pi_{\rm FC}(K=2)} -1 \simeq  $  
& 
& $-2 c$
& $- 4 c^2$
& $- \frac{25 }{3}c^3$
& $- 18 c^4$
& $- \frac{2401 }{60}c^5$
& $- \frac{4096 }{45}c^6$
& $- \frac{59049 }{280}c^7$
& $- \frac{31250 }{63}c^8$
\\
\hline
$ \frac{P_{\rm FC}(K=3|2 \le S \le s)}{\pi_{\rm FC}(K=3)} -1 \simeq  $  
& 
& 
& $-3c$
& $- \frac{15 }{2}c^2$
& $- 18 c^3$
& $- \frac{343 }{8}c^4$
& $- \frac{512 }{5}c^5$
& $- \frac{19683 }{80}c^6$
& $- \frac{12500 }{21}c^7$
\\
\hline
$ \frac{P_{\rm FC}(K=4|2 \le S \le s)}{\pi_{\rm FC}(K=4)} -1 \simeq  $  
& 
&
&
& $- 4 c$
& $- 12 c^2$
& $- \frac{98 }{3}c^3$
& $- \frac{256 }{3}c^4$
& $- \frac{2187 }{10}c^5$
& $- \frac{5000 }{9}c^6$
\\
\hline
$ \frac{P_{\rm FC}(K=5|2 \le S \le s)}{\pi_{\rm FC}(K=5)} -1 \simeq  $  
& 
&
&
&
& $-5 c $
& $- \frac{35 }{2}c^2$
& $- \frac{160 }{3}c^3$
& $- \frac{1215 }{8}c^4$
& $- \frac{1250 }{3}c^5$
\\
\hline
$ \frac{P_{\rm FC}(K=6| 2 \le S \le s)}{\pi_{\rm FC}(K=6)} -1 \simeq  $  
& 
& 
&
& 
&
& $- 6 c$
& $- 24 c^2$
& $- 81 c^3$
& $- 250 c^4$
\\
\hline
$ \frac{P_{\rm FC}(K=7| 2 \le S \le s)}{\pi_{\rm FC}(K=7)} -1 \simeq  $  
&   &   &    &  &  &
& $- 7 c$
& $- \frac{63 }{2}c^2$
& $- \frac{350 }{3}c^3$
\\
\hline
$ \frac{P_{\rm FC}(K=8| 2 \le S \le s)}{\pi_{\rm FC}(K=8)} -1 \simeq  $  
& & & & & & &
& $-8c $
& $- 40 c^2$
\\
\hline
$ \frac{P_{\rm FC}(K=9| 2 \le S \le s)}{\pi_{\rm FC}(K=9)} -1 \simeq  $  
& & & & 
& & & &
& $- 9 c$
\\
\hline
\hline
\end{tabular}
\label{table:Pkexp}
\end{table}

\clearpage
\newpage

\begin{table}
\caption{
The probabilities $P_{\rm FC}(L = \ell | L <\infty, S=s)$ that
a pair of random nodes on a random component of size $s$
in a subcritical ER network will be at a distance $\ell$ from each other
for small tree component of $s=2,3,\dots,10$ nodes.
}
\vspace{0.4in}
\begin{tabular}{| l  | r  | r | r | r |r | r | r | r | r |}
\hline \hline
& $s = 2$
     &  $s = 3$  &  $s = 4$ & $s = 5$ & $s = 6$ & $s = 7$ & $s = 8$ & $s = 9$ & $s = 10$ \\ 
\hline \hline 
$P_{\rm FC}(L=1 |  S=s)$ =
&
$1$
&
$\frac{2}{3}$
&
$\frac{ 1}{2}$
&
$\frac{ 2}{5}$
&
$\frac{1}{3}$
&
$\frac{2}{7}$
&
$\frac{1}{4}$
&
$\frac{2}{9}$
&
$\frac{1}{5}$
\\
\hline
$P_{\rm FC}(L=2 | S=s) =$
& 
&
$\frac{1}{3}$
&
$\frac{3}{8}$
&
$\frac{9}{25}$
&
$\frac{1}{3}$
&
$\frac{15}{49}$
&
$\frac{9}{32}$
&
$\frac{7}{27}$
&
$\frac{6}{25}$
\\
\hline
$P_{\rm FC}(L=3 | S=s) =$
&
&
&
$\frac{1}{8}$
&
$\frac{24}{125}$
&
$\frac{2}{9}$
&
$\frac{80}{343}$
&
$\frac{15}{64}$
&
$\frac{56}{243}$
&
$\frac{28}{125}$
\\
\hline
$P_{\rm FC}(L=4 | S=s) = $
&
&
&
&
$\frac{6}{125}$
&
$\frac{5}{54}$
&
$\frac{300}{2401}$
&
$\frac{75}{512}$
&
$\frac{350}{2187}$
&
$\frac{21}{125}$
\\
\hline
$P_{\rm FC}(L=5 | S=s)=$
&
&
&
&
&
$\frac{1}{54}$
&
$\frac{720}{16807}$
&
$\frac{135}{2048}$
&
$\frac{560}{6561}$
&
$\frac{63}{625}$
\\
\hline
$P_{\rm FC}(L=6 | S=s)=$
&
&
&
&
&
&
$\frac{120}{16807}$
&
$\frac{315}{16384}$
&
$\frac{1960}{59049}$
&
$\frac{147}{3125}$
\\
\hline
$P_{\rm FC}(L=7 | S=s) =$
&
&
&
&
&
&
&
$\frac{45}{16384}$
&
$\frac{4480}{531441}$
&
$\frac{252}{15625}$
\\
\hline
$P_{\rm FC}(L=8 | S=s)=$
&
&
&
&
&
&
&
&
$\frac{560}{531441}$
&
$\frac{567}{156250}$
\\
\hline
$P_{\rm FC}(L=9 | S=s)=$
&
&
&
&
&
&
&
&
&
$\frac{63}{156250}$
\\
\hline
\hline
${\mathbb E}[L | S=s]=$
&
$1$
&
$\frac{4}{3}$
&
$\frac{13}{8}$
&
$\frac{236}{125}$
&
$\frac{115}{54}$
&
$\frac{39572}{16807}$
&
$\frac{42037}{16384}$
&
$\frac{1469756}{531441}$
&
$\frac{461843}{156250}$
\\
\hline
${\mathbb E}[L^2 | S=s] =$
&
1
&
2
&
$\frac{25}{8}$
&
$\frac{542}{125}$
&
$\frac{101}{18}$
&
$\frac{116582}{16807}$
&
$\frac{136033}{16384}$
&
$\frac{1718890}{177147}$
&
$\frac{1739471}{156250}$
\\
\hline
${\rm Var}[L | S=s] =$
&
$0$
&
$\frac{2}{9}$
&
$\frac{31}{64}$
&
$\frac{12054}{15625}$
&
$\frac{3137}{2916}$
&
$\frac{393450490}{282475249}$
&
$\frac{461655303}{268435456}$
&
$\frac{580283161934}{282429536481}$
&
$\frac{58493387101}{24414062500}$
\\
\hline
\hline
\end{tabular}
\label{table:Pleqs}
\vspace{0.4in}
\end{table}

\begin{table}
\caption{
The probabilities $P_{\rm FC}(L = \ell | L<\infty,S \le s)$ that
a pair of random nodes on a random component of size $S \le s$
in a subcritical ER network will be at a distance $\ell$ from each other
for small tree components of $s=2,3,\dots,10$ nodes.
}
\vspace{0.4in}
\begin{tabular}{| l  | r  | r | r | r |r | }
\hline \hline
& $s = 2$
     &  $s = 3$  &  $s = 4$ & $s = 5$ & $s = 6$  \\ 
\hline \hline 
$P_{\rm FC}(L=1 | S \le s) =$ 
&
$1$
&
$\frac{1 + 2 \eta}{1 + 3 \eta}$
&
$\frac{1 + 2 \eta + 4 \eta^2}{1 + 3 \eta + 8 \eta^2}$
&
$\frac{6 + 12 \eta + 24 \eta^2 + 50 \eta^3}{6 + 18 \eta + 48 \eta^2 + 125 \eta^3}$
&
$\frac{6 + 12 \eta + 24 \eta^2 + 50 \eta^3 + 108 \eta^4}
{6 + 18 \eta + 48 \eta^2 + 125 \eta^3 + 324 \eta^4}$
\\
\hline
$P_{\rm FC}(L=2 | S \le s) =$
& 
&
$\frac{\eta}{1 + 3 \eta}$
&
$\frac{\eta + 3 \eta^2}{1 + 3 \eta + 8 \eta^2}$
&
$\frac{6 \eta + 18 \eta^2 + 45 \eta^3}{6 + 18 \eta + 48 \eta^2 + 125 \eta^3}$
&
$\frac{6 \eta + 18 \eta^2 + 45 \eta^3 + 108 \eta^4}
{6 + 18 \eta + 48 \eta^2 + 125 \eta^3 + 324 \eta^4}$
\\
\hline
$P_{\rm FC}(L=3 | S \le s) =$
&
&
&
$\frac{\eta^2}{1 + 3 \eta + 8 \eta^2}$
&
$\frac{6 \eta^2 + 24 \eta^3}{6 + 18 \eta + 48 \eta^2 + 125 \eta^3}$
&
$\frac{6 \eta^2 + 24 \eta^3 + 72 \eta^4}{6 + 18 \eta + 48 \eta^2 + 125 \eta^3 + 324 \eta^4}$
\\
\hline
$P_{\rm FC}(L=4 | S \le s) = $
&
&
&
&
$\frac{6 \eta^3}{6 + 18 \eta + 48 \eta^2 + 125 \eta^3}$
&
$\frac{6 \eta^3 + 30 \eta^4}{6 + 18 \eta + 48 \eta^2 + 125 \eta^3 + 324 \eta^4}$
\\
\hline
$P_{\rm FC}(L=5 | S \le s)=$
&
&
&
&
&
$\frac{6 \eta^4}{6 + 18 \eta + 48 \eta^2 + 125 \eta^3 + 324 \eta^4}$
\\
\hline
\hline
${\mathbb E}[L | S \le s] =$
&
$1$
&
$\frac{1 + 4 \eta}{1 + 3 \eta}$
&
$\frac{1 + 4 \eta + 13 \eta^2}{1 + 3 \eta + 8 \eta^2}$
&
$\frac{6 + 24 \eta + 78 \eta^2 + 236 \eta^3}
{6 + 18 \eta + 48 \eta^2 + 125 \eta^3}$
&
$\frac{6 + 24 \eta + 78 \eta^2 + 236 \eta^3 + 690 \eta^4}
{6 + 18 \eta + 48 \eta^2 + 125 \eta^3 + 324 \eta^4}$
\\
\hline
${\mathbb E}[L^2 | S \le s] =$
&
$1$
&
$\frac{1+ 6 \eta }{1 + 3 \eta}$
&
$\frac{1 + 6 \eta + 25 \eta^2}{1 +3 \eta + 8 \eta^2}$
&
$\frac{6 +36 \eta + 150 \eta ^2 +   542\eta^3}{6 +18 \eta +  48 \eta^2 +  125 \eta^3}$
&
$\frac{6  +36 \eta + 150 \eta^2  +542 \eta^3 +  1818 \eta^4}
{6 +18 \eta + 48 \eta^2 +125 \eta^3 +  324 \eta^4}$
\\
\hline
\hline
\end{tabular}
\label{table:Plles}
\end{table}

\begin{table}
\caption{
The leading finite size correction terms $r_{s,\ell} c^{s-\ell}$ 
of Eq. (\ref{eq:r_sell}) for the
DSPL over all the tree topologies with up to $s$ nodes.
The distribution 
$P_{\rm FC}(L=\ell | L < \infty)=(1-c)c^{\ell-1}$,
given by Eq. (\ref{eq:DSPL}),
is the DSPL over all pairs of nodes that reside on the 
same component in the entire subcritical network.
As $s$ is increased, the correction term decreases as $c^{s-2}$
and $P_{\rm FC}(L=\ell | L<\infty,S \le s)$
converges towards
$P_{\rm FC}(L=\ell | L<\infty)$.
}
\hspace{0.2in}
\begin{tabular}{| l  | r | r | r |r | r | r | r | r | r |}
\hline \hline
   & $s=2$  &  $s = 3$  &  $s = 4$ & $s = 5$ & $s = 6$ & $s = 7$ & $s = 8$ & $s = 9$ & $s = 10$ \\ 
\hline \hline 
$ \frac{P_{\rm FC}(L=1 | L<\infty, S \le s)}{P_{\rm FC}(L=1 | L<\infty)} -1 =  $
&
$1$
&
$4 c^2$
&
$\frac{25}{2} c^3$
&
$36 c^4$
&
$\frac{2401}{24} c^5$
&
$\frac{4096}{15} c^6$
&
$\frac{59049}{80} c^7$
&
$\frac{125000}{63} c^8$
&
$\frac{214358881}{40320} c^9$
\\
\hline
$ \frac{P_{\rm FC}(L=2 | L<\infty, S \le s)}{P_{\rm FC}(L=2 | L<\infty)} -1 =  $   
&
& $- 3 c $  
&  $-\frac{15}{2} c^2$ 
& $- 18 c^3$ & 
$- \frac{343}{8} c^4$  
& $- \frac{512}{5} c^5$ 
&   $- \frac{19683}{80} c^6$
& $- \frac{12500}{21} c^7$ 
& $- \frac{19487171}{13440} c^8$
\\ 
\hline  
$ \frac{P_{\rm FC}(L=3 | L<\infty, S \le s)}{P_{\rm FC}(L=3 | L<\infty)} -1 =  $   
&
& 
& $ - 4 c $   
& $ - 12 c^2 $ 
& $ - \frac{98}{3} c^3$ 
& $ - \frac{256}{3} c^4$ 
& $ - \frac{2187}{10} c^5$ 
& $ - \frac{5000}{9} c^6 $
& $ - \frac{1771561}{1260} c^7$
\\ 
\hline 
$ \frac{P_{\rm FC}(L=4 | L<\infty, S \le s)}{P_{\rm FC}(L=4 | L<\infty)} -1 =  $   &   & &
& $ - 5 c$ 
& $ - \frac{35 }{2} c^2$ 
& $ - \frac{160}{3} c^3$ 
&  $ - \frac{1215 }{8} c^4$ 
& $ - \frac{ 1250}{3} c^5$
& $ - \frac{ 161051}{144} c^6$
  \\ 
\hline 
$ \frac{P_{\rm FC}(L=5 | L<\infty, S \le s)}{P_{\rm FC}(L=5 | L<\infty)} -1 =  $   &   &    &
& 
& $ - 6 c$ 
&$ - 24  c^2$ 
&   $ - 81 c^3$
& $ - 250 c^4$ 
& $ - \frac{14641  }{20} c^5$
\\ 
\hline 
$ \frac{P_{\rm FC}(L=6 | L<\infty, S \le s)}{P_{\rm FC}(L=6 | L<\infty)} -1 =  $   &   &   & &
& 
&$ - 7 c$ 
& $ - \frac{63 }{2} c^2$ 
& $ - \frac{ 350 }{ 3 } c^3$ 
& $ - \frac{  9317 }{24} c^4$
 \\ 
\hline
$\frac{P_{\rm FC}(L=7 | L<\infty, S \le s)}{P_{\rm FC}(L=7 | L<\infty)} - 1 = $  & & & & &
& 
& $ - 8 c$ 
& $ - 40 c^2$
& $ - \frac{484}{3} c^3$
 \\
\hline
$\frac{P_{\rm FC}(L=8 | L<\infty, S \le s)}{P_{\rm FC}(L=8 | L<\infty)} - 1 = $  & & & & & &
& 
& $- 9 c$
& $ - \frac{99}{2} c^2$
 \\
\hline
$\frac{P_{\rm FC}(L=9 | L<\infty, S \le s)}{P_{\rm FC}(L=9 | L<\infty)} - 1 = $  & & & & & & & &
& $- 10 c $
\\
\hline
\hline
\end{tabular}
\label{table:Plexp}
\end{table}

\end{document}